\documentclass{egpubl}
\usepackage{eurovis2021}

\SpecialIssuePaper         % uncomment for final version of Computer Graphics Forum, special issue
\usepackage[T1]{fontenc}
\usepackage{dfadobe}  

\usepackage{cite}  % comment out for biblatex with backend=biber
\BibtexOrBiblatex
\electronicVersion
\PrintedOrElectronic
\ifpdf \usepackage[pdftex]{graphicx} \pdfcompresslevel=9
\else \usepackage[dvips]{graphicx} \fi

\usepackage{egweblnk}
\usepackage{mathptmx}
\usepackage{url}

\usepackage{amsfonts}
\usepackage{amsmath}
\usepackage{amssymb}

\usepackage{color}
\usepackage{xcolor}

\definecolor{gray}{gray}{0.5}

\graphicspath{{Figures/}{./}}

\newcommand{\SE}{{\mathcal{H}}} % Shannon entropy
\newcommand{\DKL}{{\mathcal{D}_{\text{KL}}}} % KL-divergence
\newcommand{\DJS}{{\mathcal{D}_{\text{JS}}}} % JS-divergence
\newcommand{\Dnew}{{\mathcal{D}^k_{\text{new}}}} % the new divergence by Chen and Sbert
\newcommand{\Dncm}{{\mathcal{D}^k_{\text{ncm}}}} % the new asymmetric divergence by Chen and Sbert

\newcommand{\DnewA}{{\mathcal{D}^{k=1}_{\text{new}}}}
\newcommand{\DnewB}{{\mathcal{D}^{k=2}_{\text{new}}}}
\newcommand{\DncmA}{{\mathcal{D}^{k=1}_{\text{ncm}}}}
\newcommand{\DncmB}{{\mathcal{D}^{k=2}_{\text{ncm}}}}

\usepackage[condensed]{tgheros}
\newenvironment{narrowfont}{\fontfamily{qhvc}\selectfont}{\par}

\title[A Bounded Measure for Estimating the Benefit of Visualization: Case Studies and Empirical Evaluation]%
      {A Bounded Measure for Estimating the Benefit of Visualization:\\
      Case Studies and Empirical Evaluation}

\author[Min Chen et al.]
{\parbox{\textwidth}{\centering Min Chen$^1$\orcid{0000-0001-5320-5729}, \;
    Alfie Abdul-Rahman$^2$\orcid{0000-0002-6257-876X}, \;
    Deborah Silver$^3$, \; and \;
    Mateu Sbert$^4$
    }
    \\
{\parbox{\textwidth}{\centering $^1$University of Oxford, UK, \;
    $^2$ King's College London, UK, \;
    $^3$ Rutgers University, USA, \; and \;
    $^4$ University of Girona, Spain
    } 
}
}

\begin{document}

\teaser{
  \centering
  \includegraphics[width=\linewidth]{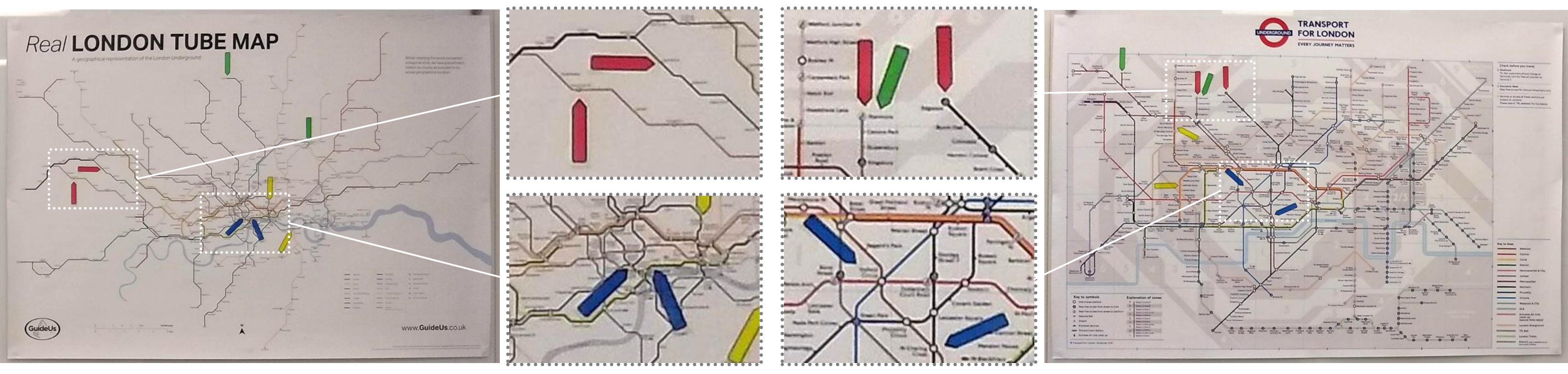}
  \caption{The London underground map (right) is a deformed map. In comparison with a relatively more faithful map (left), there is a significant amount of information loss in the deformed map, which omits some detailed variations among different connection routes between pairs of stations (e.g., distance and geometry). One common rationale is that the deformed map was designed for certain visualization tasks, which likely excluded the task for estimating the walking time between a pair of stations indicated by a pair of red or blue arrows. 
  In one of our experiments, when asked to perform such tasks using the deformed map, some participants did rather well. Can information theory explain this phenomenon? Can we quantitatively measure some relevant factors in this visualization process?}
  \label{fig:InfoLoss}
}

\maketitle
%-------------------------------------------------------------------------
\begin{abstract}%
Many visual representations, such as volume-rendered images and metro maps, feature a noticeable amount of information loss. At a glance, there seem to be numerous opportunities for viewers to misinterpret the data being visualized, hence undermining the benefits of these visual representations. In practice, there is little doubt that these visual representations are useful. The recently-proposed information-theoretic measure for analyzing the cost-benefit ratio of visualization processes can explain such usefulness experienced in practice, and postulate that the viewers’ knowledge can reduce the potential distortion (e.g., misinterpretation) due to information loss. This suggests that viewers’ knowledge can be estimated by comparing the potential distortion without any knowledge and the actual distortion with some knowledge. In this paper, we describe several case studies for collecting instances that can (i) support the evaluation of several candidate measures for estimating the potential distortion distortion in visualization, and (ii) demonstrate their applicability in practical scenarios. Because the theoretical discourse on choosing an appropriate bounded measure for estimating the potential distortion is yet conclusive, it is the real world data about visualization further informs the selection of a bounded measure, providing practical evidence to aid a theoretical conclusion. Meanwhile, once we can measure the potential distortion in a bounded manner, we can interpret the numerical values characterizing the benefit of visualization more intuitively. 
%-------------------------------------------------------------------------
%  ACM CCS 1998
%  (see https://www.acm.org/publications/computing-classification-system/1998)
% \begin{classification} % according to https://www.acm.org/publications/computing-classification-system/1998
% \CCScat{Computer Graphics}{I.3.3}{Picture/Image Generation}{Line and curve generation}
% \end{classification}
%-------------------------------------------------------------------------
%  ACM CCS 2012
%   (see https://www.acm.org/publications/class-2012)
%The tool at \url{http://dl.acm.org/ccs.cfm} can be used to generate
% CCS codes.
%Example:
% <concept_id>10010147.10010371.10010352.10010381</concept_id>
% <concept_desc>Computing methodologies~Collision detection</concept_desc>
% <concept_significance>300</concept_significance>
% <concept_id>10010583.10010588.10010559</concept_id>
% <concept_desc>Hardware~Sensors and actuators</concept_desc>
% <concept_significance>300</concept_significance>
% <concept_id>10010583.10010584.10010587</concept_id>
% <concept_desc>Hardware~PCB design and layout</concept_desc>
% <concept_significance>100</concept_significance>

% \ccsdesc[300]{Computing methodologies~Collision detection}
% \ccsdesc[300]{Hardware~Sensors and actuators}
% \ccsdesc[100]{Hardware~PCB design and layout}

\printccsdesc   
\end{abstract}


\section{Introduction}
\label{sec:Introduction}

This paper is concerned with the measurement of the benefit of visualization and viewers' knowledge used in visualization.
The history of measurement science shows that the development of measurements in different fields has not only stimulated scientific and technological advancements, but also encountered some serious contentions due to instrumental, operational, and social conventions \cite{Klein:2012:book}.
While the development of measurement systems, methods, and standards for visualization may take decades of research, one can easily imagine their impact to visualization as a scientific and technological subject.

``Measurement ... is defined as the assignment of numerals to objects or events according to rules'' \cite{Stevens:1946:S}.
Rules may be defined based on physical laws (e.g., absolute zero temperature), observational instances (e.g., the freezing and boiling points of water), or social traditions (e.g., seven days per week).
Without exception, measurement development in visualization aims to discover and define rules that will enable us to use mathematics in describing, differentiating, and explaining phenomena in visualization, as well as predicting the impact of a design decision, diagnosing shortcomings in visual analytics workflows, and formulating solutions for improvement.

In a separate and related paper (see the supplementary materials), Chen and Sbert examined a number of candidate measures for assigning numerals to the notion of potential distortion, which is one of the three components in the information-theoretic measure for quantifying the cost-benefit of visualization \cite{Chen:2016:TVCG}. They used several ``conceptual rules'' to evaluate these candidates, narrowing them down to five candidates. In this work, we focus on the remaining five candidate measures and evaluate them based on empirical evidence.
We use two synthetic case studies and two experimental case studies to instantiate values that may be returned by the candidates. 
The main ``observational rules'' used in this work include:
\begin{itemize}
    \item[a.] Does the numerical ordering of observed instances match with the intuitively-expected ordering?
    \item[b.] Does the gap between positive and negative values indicate a meaningful critical point (i.e., zero benefit in our case)? 
\end{itemize}

\noindent Although one might consider observational rules are subjective and thus undesirable, we can appreciate their roles in the development of many measurement systems used today (e.g., number of months per year, and sign of temperature in Celsius). It would be hasty to dismiss such rules, especially at this early stage of measurement development in visualization.

In addition, we use the data collected in two visualization case studies to explore the relationship between the benefit of visualization and the viewers' knowledge used in visualization.
As shown in Figure \ref{fig:InfoLoss}, in one case study, we asked participants to perform tasks for estimating the walking time (in minutes) between two underground stations indicated by a pair of red or blue arrows.
Although the deformed London underground map was not designed to perform visualization tasks, many participants performed rather well, including those that had very limited experience of using the London underground.
We use different candidate measures to estimate the supposed potential distortion.
When the captured data shows that the amount of actual distortion is lower than the supposed distortion, this indicate that the viewers' knowledge has been used in the visualization process to alleviate the potential distortion.
While we use the experiments to collect practical instances to evaluate the candidate measures empirically, they also demonstrate that we are getting closer to be able to estimate the ``benefit'' and ``potential distortion'' of practical visualization processes.

Readers are encouraged to consult the preceding paper in the supplementary materials for information about the mathematical background, the formulation of some new candidate measures, and the conceptual evaluation of the candidate measures. Nevertheless, this paper is written in a self-contained manner. 

% ====================
\section{Related Work}
\label{sec:RelatedWork}
This paper and its preceding paper (in the supplementary materials) are concerned with information-theoretic measures for quantifying aspects of visualization, such as benefit, knowledge, and potential misinterpretation.
The preceding paper focuses its review on previous information-theoretic work in visualization. In this section, we focus our review on previous measurement work in visualization.

\noindent\textbf{Measurement Science.} \quad
There is currently no standard method for measuring the benefit of visualization, levels of visual abstraction, the human knowledge used in visualization, or the potential to misinterpret visual abstraction. While these are considered to be complex undertakings, many scientists in the history of measurement science would have encountered similar challenges \cite{Klein:2012:book}.

In their book \cite{Boslaugh:2008:book}, Boslaugh and Watters describe measurement as ``the process of systematically assigning numbers to objects and their properties, to facilitate the use of mathematics in studying and describing objects and their relationships.'' They emphasize in particular measurement is not limited to physical qualities such as height and weight, but also abstract properties such as intelligence and aptitude. Pedhazur and Schmelkin \cite{Pedhazur:1991:book} assert the necessity of an integrated approach for measurement development, involving data collection, mathematical reasoning, technology innovation, and device engineering. Tal \cite{Tal:2015:book} points out that measurement is often not totally ``real'', involves the representation of ideal systems, and reflects conceptual, metaphysical, semantic, and epistemological understandings.  
Schlaudt \cite{Schlaudt:2020:book} goes one step further, referring measurement as a cultural technique.

This work is particularly inspired by the historical development of temperature scales and seismic magnitude. The former attracted the attention of many well-known scientists, benefited from both experimental observations (e.g., by Celsius, Delisle, Fahrenheit, Newton, etc.) and theoretical discoveries (e.g., by Boltzmann, Thomson (Kelvin), etc.).
The latter started not long ago as the Richter scale was outlined in 1935, and since then there have been many schemes proposed relating different physical properties.
Many scales in both applications are related to some logarithmic transformations one way or another.

\noindent\textbf{Metrics Development in Visualization.} \quad
Behrisch et al. \cite{Behrisch:2018:STAR} presented an extensive survey of quality metrics for information visualization.
Bertini et al. \cite{Bertini:2011:TVCG} described a systemic approach of using quality metrics for evaluating visualization in high-dimensional data visualization focusing on scatter plots and parallel coordinates plots.
A variety of quality metrics have been proposed to measure many different attributes, such as
abstraction quality \cite{Tufte:1986:Book, Cui:2006:TVCG, Johansson:2008:CGF},
quality of scatter plots \cite{Friedman:1974:Trans, tukey1985, Bertini:2004:SmartGraphics, wilkinson2005, sips:2009:CGF, tatu2009, tatu:2010:VQM},
quality of parallel coordinates plots \cite{Dasgupta:InfoVis:2010},
cluttering \cite{Peng2004, rosenholtz2007, yu:jov:2014},
aesthetics \cite{Filonik:2009:IV},
visual saliency \cite{Janicke:2010:CGF}, and
color mapping \cite{Bernard:2015:SPIE, Mittelstadt:2015:EuroVis, Mittelstadt:2015:CGF, Gramazio:2017:TVCG}.
In particular, J\"{a}nicke et al. \cite{Janicke:2010:CGF}
first considered a metric for estimating the amount of original data that is depicted by visualization and may be reconstructed by viewers.
Chen and Golan \cite{Chen:2016:TVCG} used the abstract form of this idea in defining their cost-benefit ratio.
While the work by J\"{a}nicke et al. \cite{Janicke:2010:CGF} relied computer vision techniques to reconstruction, this work focused on collecting and analysing empirical data because human knowledge has a major role to play in information reconstruction.

% \textcolor{blue}{Review on empirical studies that involving stimuli with ``information loss'' such as lower visual resolution than data resolution, distortion in map visualization, non-photorealistic visualization.}
% \vspace{3cm}

\noindent\textbf{Measurement in Empirical Experiments.} \quad
Almost all empirical studies in visualization involve measuring participants' performance in visualization processes. Such measured data allow us to assess the benefit of visualization or potential misinterpretation, typically in terms of accuracy and response time. Many uncontrolled empirical studies also collect participants' experience and opinions qualitatively.
The empirical studies related to the key components are those on the topics of visual abstraction and human knowledge in visualization.
Isenberg \cite{Isenberg2013} presented a survey of evaluation techniques on non-photorealistic and illustrative rendering. 
Isenberg et al. \cite{Isenberg:2006:NPAR} reported an observational study comparing hand-drawn and computer-generated non-photorealistic rendering.
Cole et al. \cite{Cole:2009:SIGGRAPH} performed a study evaluating the effectiveness of line drawing in representing shape. Mandryk et al.
\cite{Mandryk:2011:NPA} evaluated the emotional responses to non-photorealistic generated images.
Liu and Li \cite{liu:2016:ISPRS} presented an eye-tracking study examining the effectiveness and efficiency of schematic design of 30$^{\circ}$ and 60$^{\circ}$ directions in underground maps.
Hong et al. \cite{Hong:2018:CHI} evaluated the usefulness of distance cartograms map ``in the wild''.
These studies confirmed that visualization users can deal with significant information loss due to visual abstraction in many situations.

Tam et al. \cite{Tam:2017:TVCG} reported an observational study, and discovered that machine learning developers entered a huge amount of knowledge (measured in bits) into a visualization-assisted model development process.
Kijmongkolchai et al. \cite{Kijmongkolchai:2017:CGF} reported a study design for detecting and measuring human knowledge used in visualization, and translated the traditional accuracy values to information-theoretic measures.
They encountered some undesirable properties of the Kullback-Leibler divergence in their calculations.
In this work, we collect empirical data to evaluate the mathematical solutions proposed to address the issue encountered in \cite{Kijmongkolchai:2017:CGF}.

% ====================================================
\section{Overview, Notations, and Problem Statement}
\label{sec:Overview}
\noindent\textbf{Brief Overview.} \quad
Whilst hardly anyone in the visualization community would support any practice intended to deceive viewers, there have been many visualization techniques that inherently cause distortion to the original data.
The deformed London underground map in Figure \ref{fig:InfoLoss} shows such an example.
The distortion in this example is largely caused by many-to-one mappings.
A group of lines that would be shown in different lengths in a faithful map are now shown with the same length.
Another group of lines that would be shown with different geometric shapes are now shown as the same straight line.
In terms of information theory, when the faithful map is transformed to the deformed, a good portion of information has been lost because of these many-to-one mappings.

The common phrase that ``the appropriateness of information loss depends on tasks'' is not an invalid explanation. Partly by a similar conundrum in economics ``what is the most appropriate resolution of time series for an economist'', Chen and Golan proposed an information-theoretic cost-benefit ratio for measuring various factors involved in visualization processes \cite{Chen:2016:TVCG}. Its qualitative version is:
\begin{equation} \label{eq:CBM-1}
    \frac{\text{Benefit}}{\text{Cost}} = \frac{\text{Alphabet Compression} - \text{Potential Distortion}}{\text{Cost}}
\end{equation}

Appendix \ref{app:OriginalTheory} provides more detailed explanation of this measure, while Appendix \ref{app:TasksUsers} explains in detail how tasks and users are considered by this measure in abstract.

\begin{figure}[t]
\centering
\includegraphics[width=\linewidth]{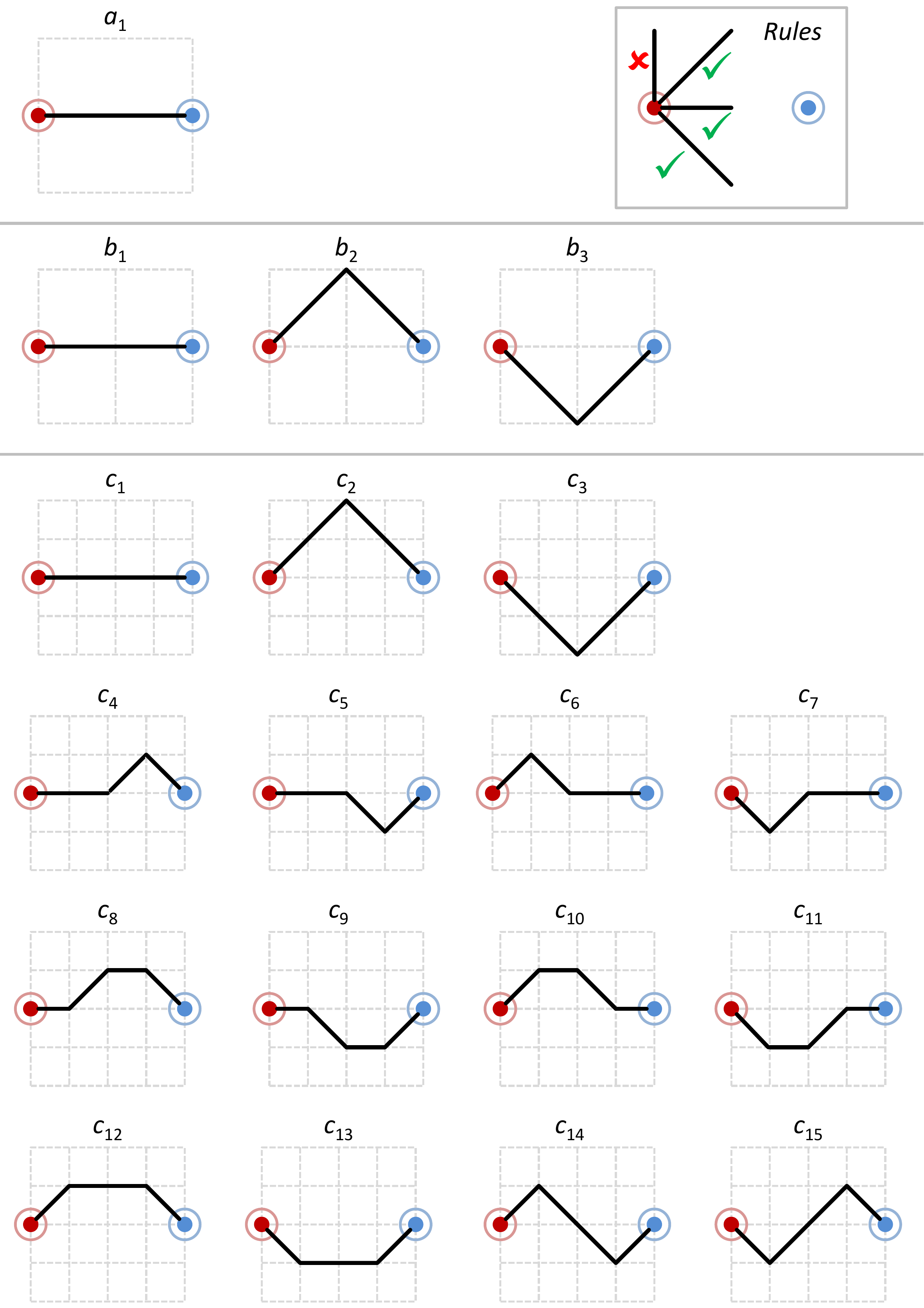}
\caption{Three alphabets illustrate possible metro maps (letters) in different grid resolutions.}
\label{fig:Math}
\vspace{-4mm}
\end{figure}

\noindent\textbf{Mathematical Notations.} \quad
Consider a simple metro map consists of only two stations in Figure \ref{fig:Math}.
We consider three different grid resolutions, with $1 \times 1$ cell, $2 \times 2$ cells, and $4 \times 4$ cells respectively.
The following set of rules determine whether a potential path is allowed or not:
\begin{itemize}
    \item The positions of the two stations are fixed on each of the three grids and there is only one path between the red station and the blue station;
    \item As shown on the top-right of Figure \ref{fig:Math}, only horizontal, and diagonal path-lines are allowed;
    \item When one path-line joins another, it can rotate by up to $\pm 45^\circ$;
    \item All joints of path-lines can only be placed on grid points;
\end{itemize}

For the first grid with $1 \times 1$ cell, there is only one possible path.
We define an alphabet $\mathbb{A}$ to contain this option as its only letter $a_1$, i.e., $\mathbb{A} = \{ a_1 \}$.
For the second grid with $2 \times 2$ cells, we have an alphabet $\mathbb{B} = \{b_1, b_2, b_3\}$, consisting of three optional paths.
For the third grid with $4 \times 4$ cells, there are 15 optional paths, which are letters of alphabet $\mathbb{C} = \{c_1, c_2, \ldots, c_{15}\}$.
When the resolution of the grid increases, the alphabet of options becomes bigger quickly. We can imagine it gradually allows the designer to create a more faithful map.
To ensure the calculation below is easy to follow, we consider only the first two grids below. 

To a designer of the underground map, at the $1 \times 1$ resolution, there is only one choice regardless how much the designer would like to draw the path to reflect the actual geographical path of the metro line between these two stations. At the $4 \times 4$ resolution, the designer has many options.
Hence there is more uncertainty associated with the third grid.
This uncertainty can be measured by Shannon entropy, which is defined as:
\[
    \SE(\mathbb{Z}) = - \sum_{i=1}^n p_i \log_2 p_i \quad \text{where } p_i \in [0, 1], \sum_{i=1}^n p_i = 1
\]
where $\mathbb{Z}$ is an alphabet, and can be replaced with $\mathbb{A}$, $\mathbb{B}$, or  $\mathbb{C}$. To calculate Shannon entropy, the alphabet $\mathbb{Z}$ needs to be accompanied by a \emph{probability mass function} (PMF), which is written as $P(\mathbb{Z})$.
Each letter $z_i \in \mathbb{Z}$ is thus associated with a probability value $p_i \in P$.

\noindent Note: In this paper, to simplify the notations in different contexts, for an information-theoretic measure, we use an alphabet $\mathbb{Z}$ and its
PMF $P$ interchangeably, e.g., $\SE(P(\mathbb{Z})) = \SE(P) = \SE(\mathbb{Z})$. Appendix
\ref{app:InfoTheory} provides more mathematical background about information theory, which may be helpful to some readers.

Let first consider the single-letter alphabet $\mathbb{A}$ and its PMF $Q$.
Because $n = 1$ and $q_1 = 1$, we have $\SE(\mathbb{A}) = 0$ bits.
The alphabet is 100\% certain, reflecting the fact that the designer has no choice.

The alphabet $\mathbb{B}$ has three design options $b_1$, $b_2$, and $b_3$. If they have an equal chance to be selected by the designer, we have a PMF $Q_u$ with $q_1 = q_2 = q_3 = 1/3$, and thus $\SE(Q_u(\mathbb{B})) \approx 1.585$ bits.
When we examine the three options in Figure \ref{fig:Math}, it is not unreasonable to consider a second scenario that the choice may be in favour of the straight line option $b_1$ in designing a metro map according to the real geographical data.
If a different PMF $Q_v$ is given as $q_1 = 0.9, q_2 = q_3 = 0.05$, we have $\SE(Q_v(\mathbb{B})) \approx 0.569$ bits.
The second scenario features less entropy and is thus of more certainty.

Consider that the designer is given a metro map designed using alphabet $\mathbb{B}$, and is asked to produce a more abstract map using alphabet $\mathbb{A}$.
To the designer, it is a straightforward task, since there is only one option in $\mathbb{A}$.
When a group of viewers are visualizing the final design $a_1$, we could give these viewers a task to guess what may be the original map designed with $\mathbb{B}$.
If most viewers have no knowledge about the possible options $b_2$ and $b_3$, and almost all choose $b_1$ as the original design, we can describe their decisions using a PMF $P$ such that $p_1 = 0.998, p_2 = p_3 = 0.001$.
Since $P$ is not the same as either $Q_u$ or $Q_v$, the viewers' decisions diverge from the actual PMF associated with $\mathbb{B}$. This divergence can be measured using KL-divergence:
\[
    \DKL(P(\mathbb{Z})||Q(\mathbb{Z})) = \sum_{i=1}^n p_i (\log_2 p_i - \log_2 q_i)
    =  \sum_{i=1}^n p_i \log_2 \frac{p_i}{q_i}
\]

Using $\DKL$, we can calculate (i) if the original design alphabet $\mathbb{B}$ has the PMF $Q_u$, we have $\DKL(P||Q_u) \approx 1.562$ bits; and (ii) if the original design alphabet $\mathbb{B}$ has the PMF $Q_v$, we have $\DKL(P||Q_v) \approx 0.138$ bits.
There is more divergence in the case (i) than case (ii).
Intuitively we can guess this as $P$ appears to be similar to $Q_v$.

Recall the qualitative formula in Eq.\,\ref{eq:CBM-1}.
In \cite{Chen:2016:TVCG}, the benefit of a visual analytics process is defined as:
\begin{equation} \label{eq:CBM-2}
  \text{Benefit} = \text{AC} - \text{PD}
                 = \SE(\mathbb{Z}_i) - \SE(\mathbb{Z}_{i+1})
                 - \DKL(\mathbb{Z}'_i||\mathbb{Z}_i)
\end{equation}
\noindent where $\mathbb{Z}_i$ is the input alphabet to the process and $\mathbb{Z}_{i+1}$ is the output alphabet. $\mathbb{Z}'_i$ is an alphabet reconstructed based on $\mathbb{Z}_{i+1}$. $\mathbb{Z}'_i$ has the same set of letters as $\mathbb{Z}_i$ but likely a different PMF.

% For the example of simple metro maps depicted in Figure \ref{fig:Math}, we can now relate the designer's process and viewers' process back to the workflow in Figure \ref{fig:Concept}. We can associate alphabet $\mathbb{B}$ with the block ``Data'', alphabet $\mathbb{A}$ with the block showing a deformed metro map.
In terms of Eq.\,\ref{eq:CBM-2}, we have
$\mathbb{Z}_{i} = \mathbb{B}$ with PMF $Q_u$ or $Q_v$, 
$\mathbb{Z}_{i+1} = \mathbb{A}$ with PMF $Q$, and
$\mathbb{Z}'_{i} = \mathbb{B}'$ with PMF $P$.
We can thus calculate the benefit in the two cases as:
\begin{align*}
    \text{Benefit of case (i)} &= \SE(\mathbb{B}) - \SE(\mathbb{A}) - \DKL(\mathbb{B}'||\mathbb{B})\\
    &= \SE(Q_u) - \SE(Q) - \DKL(P||Q_u)\\
    &\approx 1.585 - 0 - 1.562 = 0.023 \text{bits}
\end{align*}
\begin{align*}
    \text{Benefit of case (ii)} &= \SE(Q_v) - \SE(Q) - \DKL(P||Q_v)\\
    &\approx 0.569 - 0 - 0.138 = 0.431 \text{bits}
\end{align*}
In the case (ii), because the viewers' expectation is closer to the original PMF $Q_v$, there is more benefit in the visualization process than the case (i) though the case (ii) has less AC than case (i). 

However, $\DKL$ has an undesirable mathematical property.
If we consider a third case, (iii), where the original PMF $Q_w$ is strongly in favour of $b_2$, such as $q_1 = \epsilon, q_2 = 1-\epsilon, q_3 = \epsilon$, where $0 < \epsilon < 1$ is a small positive value.
If $\epsilon = 0.001$, $\DKL(P||Q_w) = 9.933$ bits.
If $\epsilon \rightarrow 0$, $\DKL(P||Q_w) \rightarrow \infty$.
Since the maximum entropy (uncertainty) for $\mathbb{B}$ is only about 1.585, it is difficult to interpret that viewers' divergence can be more than that maximum, not mentioning the infinity. 

% ----------------------------------------
\begin{table*}[t]
  \centering
  \caption{A summary of multi-criteria analysis. Each measure is scored against a criterion using an integer in [0, 5] with 5 being the best.}
  \label{tab:MultiCriteria}
  \begin{tabular}{@{}l@{\hspace{3mm}}c@{\hspace{3mm}}c@{\hspace{3mm}}c@{\hspace{2mm}}c@{\hspace{2mm}}c@{\hspace{3mm}}c@{\hspace{3mm}}c@{\hspace{3mm}}c@{\hspace{3mm}}c@{\hspace{3mm}}c@{}}
  \textbf{Criteria} & \textbf{Importance}
  & $0.3\DKL$ & $\DJS$ & $\SE(P|Q)$
  & $\DnewA$ & $\DnewB$
  & $\DncmA$ & $\DncmB$
  & $D^{k=2}_{\text{M}}$ & $D^{k=200}_{\text{M}}$\\[0.5mm]
  \hline
  1. Boundedness & critical
        & 0 & 5 & 5 & 5 & 5 & 5 & 5 & 3 & 3 \\
        \multicolumn{11}{l}{$\blacktriangleright$
        \emph{$0.3\DKL$ is eliminated but used below only for comparison.
        The other scores are carried forward.}} \\
  \hline
  2. Number of PMFs & important
        & \textcolor{gray}{5} & 5 & 2 & 5 & 5 & 5 & 5 & 5 & 5\\ 
  3. Entropic measures & important
        & \textcolor{gray}{5} & 5 & 5 & 5 & 5 & 5 & 5 & 1 & 1 \\
  4. Curve shapes & helpful
        & \textcolor{gray}{5} & 5 & 1 & 2 & 4 & 2 & 4 & 3 & 3 \\
  5. Curve shapes & helpful
        & \textcolor{gray}{5} & 4 & 1 & 3 & 5 & 3 & 5 & 2 & 3\\[0.5mm]
  \multicolumn{2}{l}{$\blacktriangleright$
        \emph{Eliminate} \emph{$\SE(P|Q)$}, $\mathcal{D}^2_{\text{M}}$, $\mathcal{D}^{200}_{\text{M}}$ \emph{based on criteria 1-5}}
        & \textbf{sum:} & \textbf{24} & \textcolor{gray}{\textbf{14}}
        & \textbf{20} & \textbf{24} & \textbf{20} & \textbf{24}
        & \textcolor{gray}{\textbf{14}} & \textcolor{gray}{\textbf{15}}\\[0.5mm]
  \hline
  6. Scenario: \emph{good} and \emph{bad} (Figure \ref{fig:Compare-GoodBad}) & helpful
        & $-$ & 3 & $-$ & 5 & 4 & 5 & 4 & $-$ & $-$\\
  7. Scenario: A, B, C, D (Figure \ref{fig:Compare-ABCD}) & helpful
        & $-$ & 4 & $-$ & 5 & 3 & 2 & 1 & $-$ & $-$\\
  8. Case Study 1 (Section \ref{sec:VolVis}) & important
        & $-$ & 5 & $-$ & 1 & 5 & 5 & 5 & $-$ & $-$\\
  9. Case Study 2: (Section \ref{sec:London}) & important
        & $-$ & 3 & $-$ & 1 & 5 & 3 & 3 & $-$ & $-$\\[0.5mm]
        \multicolumn{2}{l}{$\blacktriangleright$
        \emph{$\DnewB$ has the highest score based on criteria 6-9 (1-9)}}
        & \textbf{sum:} & \textcolor{gray}{\textbf{15}(39)} &
        & \textcolor{gray}{\textbf{12}(32)} & \textbf{17}(41)
        & \textcolor{gray}{\textbf{15}(35)} & \textcolor{gray}{\textbf{13}(37)}\\[0.5mm]
   \hline
  \end{tabular}
  \vspace{-0mm}
\end{table*}

% ----------------------------------------
\noindent\textbf{Problem Statement.} \quad
When using $\DKL$ in Eq.\, \ref{eq:CBM-1} in a relative or qualitative context (e.g., \cite{Chen:2019:TVCG,Chen:2019:CGF}), the unboundedness of the KL-divergence does not pose an issue.
However, this does become an issue when the KL-divergence is used to measure PD in an absolute and quantitative context.

In the preceding paper (see the supplementary materials), Chen and Sbert showed that conceptually, it is the unboundedness is not consistent with a conceptual interpretation of KL-divergence for measuring the inefficiency of a code (alphabet) that has a finite number of codewords (letters).
They propose to find a suitable bounded divergence measure to replace the $\DKL$ term.
They examined eight candidate measures, analysed their mathematical properties with the aid of visualization, and narrowed down to five measures using multi-criteria decision analysis (MCDA) \cite{Ishizaka:2013:book}.
The upper half of Table \ref{tab:MultiCriteria} shows their conceptual evaluation based on five criteria.
In this work, we continue their MCDA process by introducing criteria based on the analysis of instances obtained when using the remaining five candidate measures in different case studies, which correspond to criteria 6 -- 9 respectively in Table \ref{tab:MultiCriteria}.

For self-containment, we give the mathematical definition of the five candidate measures below. In this paper, we treat them as black-box functions, since they have already undergone the conceptual evaluation in the preceding paper. For more detailed conceptual and mathematical discourse on these five candidate measures, please consult the preceding paper in the supplementary materials. 

The first candidate measure is Jensen-Shannon divergence \cite{Lin:1991:TIT}.
The conceptual evaluation yield a promising score of 24 in Table \ref{tab:MultiCriteria}.
It is defined as:
\begin{equation} \label{eq:DJS}
\begin{split}
    \DJS(P||Q) &= \frac{1}{2} \bigl( \DKL(P||M) + \DKL(Q||M) \bigr) = \DJS(Q||P)\\
    &= \frac{1}{2} \sum_{i=1}^n \biggl(p_i \log_2 \frac{2 p_i}{p_i + q_i} + q_i \log_2 \frac{2 q_i}{p_i + q_i} \biggr)
\end{split}
\end{equation}
\noindent where $P$ and $Q$ are two PMFs associated with the same alphabet $\mathbb{Z}$ and $M$ is the average distribution of $P$ and $Q$.
Each letter $z_i \in \mathbb{Z}$ is associated with a probability value $p_i \in P$ and another $q_i \in Q$. 
With the base 2 logarithm as in Eq.\,\ref{eq:DJS}, $\DJS(P||Q)$ is bounded by 0 and 1.

The second and third candidate measures are two instances of a new measure $\Dnew$ proposed by Chen and Sbert in the preceding work. The two instances are $\Dnew (k=1)$ and $\Dnew (k=2)$.
They received scores of 20 and 24 respectively in the conceptual evaluation.
$\Dnew$ is defined as follows:
\begin{equation} \label{eq:New}
    \Dnew(P||Q) = \frac{1}{2} \sum_{i=1}^n (p_i + q_i) \log_2 \bigl( |p_i - q_i|^k + 1 \bigr)
\end{equation}
\noindent where $k>0$. Because $0 \leq |p_i - q_i|^k \leq 1$, we have
\[
    \frac{1}{2} \sum_{i=1}^n (p_i + q_i) \log_2 (0+1)
    \leq \Dnew(P||Q) \leq \frac{1}{2} \sum_{i=1}^n (p_i + q_i) \log_2 (1+1)
\]
Since $\log_2 1=0$, $\log_2 2=1$, $\sum p_i = 1$, $\sum q_i = 1$, $\Dnew(P||Q)$ is thus bounded by 0 and 1.

The fourth and five candidate measures are two instances of a non-commutative version of $\Dnew$. It is denoted as $\Dncm$, and the two instances are $\Dncm (k=1)$ and $\Dncm (k=2)$.
They also received scores of 20 and 24 respectively in the conceptual evaluation.
$\Dncm$ is defined as follows:
\begin{equation} \label{eq:NewA}
    \Dncm(P||Q) = \sum_{i=1}^n p_i \log_2 \bigl( |p_i - q_i|^k + 1 \bigr)
\end{equation}
which captures the non-commutative property of $\DKL$.

As $\DJS$, $\Dnew$, and $\Dncm$ are bounded by [0, 1], if any of them is selected to replace $\DKL$, Eq.\,\ref{eq:CBM-2} can be rewritten as 
\begin{equation} \label{eq:CBM-3}
  \text{Benefit} = \SE(\mathbb{Z}_i) - \SE(\mathbb{Z}_{i+1})
                 - \SE_{\text{max}}(\mathbb{Z}_i) \mathcal{D}(\mathbb{Z}'_i||\mathbb{Z}_i)
\end{equation}
\noindent where $\SE_{\text{max}}$ denotes maximum entropy, while $\mathcal{D}$ is a placeholder for $\DJS$, $\Dnew$, or $\Dncm$.

\begin{figure*}[t]
    \centering
    \includegraphics[width=178mm]{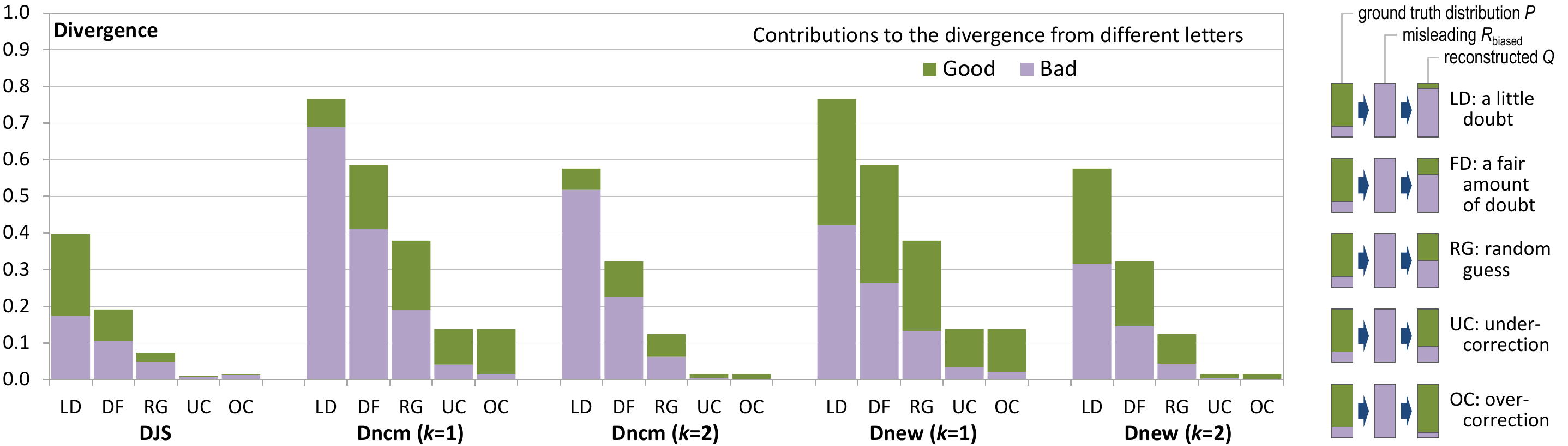}
    \caption{An example scenario with two states \emph{good} and \emph{bad} has a ground truth PMF $P=\{0.8, 0.2\}$. From the output of a biased process that always informs users that the situation is \emph{bad}. Five users, LD, DF, RG, UC, and OC, have different knowledge, and thus different divergence. The five candidate measures return different values of divergence. We would like to see which set of values are more intuitive. The illustration on the right shows two transformations of the alphabets and their PMFs, one by the misleading communication and the other by the reconstruction.}
    \label{fig:Compare-GoodBad}
    \vspace{-2mm}
\end{figure*}

\begin{figure*}[t]
    \centering
    \includegraphics[width=178mm]{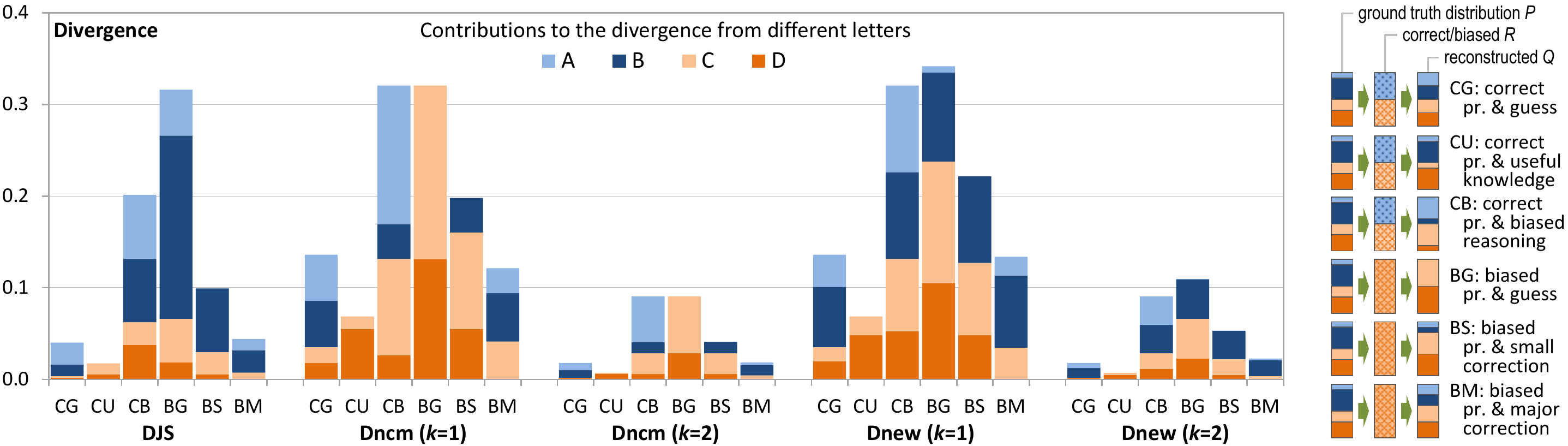}
    \caption{An example scenario with four data values: A, B, C, and D. Two processes (one correct and one biased) aggregated them to two values AB and CD. Users CG, CU, CB attempt to reconstruct [A, B, C, D] from the output [AB, CD] of the correct process, while BG, BS, and BM attempt to do so with the output from the biased processes.
    The bar chart shows divergence values of the six users computed using the five candidate measures. The illustration on the right shows two transformations of the alphabets and their PMFs, one by the correct or biased process (pr.) and the other by the reconstruction.
    }
    \label{fig:Compare-ABCD}
    \vspace{-4mm}
\end{figure*}

% ==============================
\section{Synthetic Case Studies}
\label{sec:Synthetic}
The work in this section and the next section is inspired by the historical development of temperature scales, in which many pioneering scientists collected observational data to help reason about the candidate scales.
We first consider two synthetic case studies, which allow us to define idealized situations, from which collected data do not contain any noise.
We use these case studies to see if the values returned by different divergence measures make sense.
In many ways, this is similar to testing a piece of software using pre-defined test cases.
Nevertheless, these test cases feature more complex alphabets than those considered by the conceptual evaluation reported in the preceding paper.
In our analysis, we followed the best-worst method of MCDA \cite{Rezaei:2015:O}.
Our main ``observational rules'' are:
\begin{itemize}
    \item[a.] Does the numerical ordering of observed instances match with the intuitively-expected ordering?
    \item[b.] Does the gap between positive and negative values indicate a meaningful critical point (i.e., zero benefit in our case)? 
\end{itemize}

\noindent\textbf{Criterion 6.} \quad
Let $\mathbb{Z}$ be an alphabet with two letters, \emph{good} and \emph{bad}, for describing a scenario (e.g., an object or an event), which has the probability of \emph{good} is $p_1 = 0.8$, and that of \emph{bad} is $p_2 = 0.2$.
In other words, $P = \{0.8, 0.2\}$.
Imagine that a biased process (e.g., a distorted visualization, an incorrect algorithm, or a misleading communication) conveys the information about the scenario always \emph{bad}, i.e., a PMF $R_{\text{biased}} = \{0, 1\}$.
Users at the receiving end of the process may have different knowledge about the actual scenario, and they will make a decision after receiving the output of the process.
For example, there are five users and we have obtained the probability of their decisions as follows:
\begin{itemize}
  % \vspace{-2mm}
  \item LD --- The user has a little doubt about the output of the process, and decides \emph{bad} 90\% of the time, and \emph{good} 10\% of the time, i.e., with PMF $Q = \{0.1, 0.9\}$.
  % \vspace{-2mm}
  \item FD --- The user has a fair amount of doubt, with $Q = \{0.3, 0.7\}$.
  % \vspace{-2mm}
  \item RG --- The user makes a random guess, with $Q = \{0.5, 0.5\}$.
  % \vspace{-2mm}
  \item UC --- The user has adequate knowledge about $P$, but under-compensate it slightly, with $Q = \{0.7, 0.3\}$.
  % \vspace{-2mm}
  \item OC --- The user has adequate knowledge about $P$, but over-compensate it slightly, with $Q = \{0.9, 0.1\}$.
\end{itemize}
We can use different candidate measures to compute the divergence between $P$ and $Q$.
Figure \ref{fig:Compare-GoodBad} shows different divergence values returned by these measures, while the transformations from $P$ to $R_\text{biased}$ and then to $Q$ are illustrated on the right margin of the figure.
Each value is decomposed into two parts, one for \emph{good} and one for \emph{bad}. 
All these measures can order these five users reasonably well.
The users UC (under-compensate) and OC (over-compensate) have the same values with $\Dnew$ and $\Dncm$, while $\DJS$ considers OC has slightly more divergence than UC (0.014 vs. 0.010).
$\DJS$ returns relatively low values than other measures.
For UC and OC, $\DJS$, $\DncmA$, and $\DnewB$ return small values $(<0.02)$, which are a bit difficult to estimate.   

$\DncmA$ and $\DncmB$ show strong asymmetric patterns between \emph{good} and \emph{bad}, reflecting the probability values in $Q$.
In other words, the more decisions on \emph{good}, the more \emph{good}-related divergence.
This asymmetric pattern is not in anyway incorrect, as the KL-divergence is also non-commutative and would also produce much stronger asymmetric patterns.
An argument for supporting commutative measures would point out that the higher probability of \emph{good} in $P$ should also influence the balance between the \emph{good}-related divergence.

We decide to score $\DJS$ 3 because of its lower valuation and its non-equal comparison of OU and OC.
We score $\DncmA$ and $\DnewA$ 5; and $\DncmB$ and $\DnewB$ 4 as the values returned by $\DncmA$ and $\DnewA$ are slightly more intuitive.

\noindent\textbf{Criterion 7.} \quad
We now consider a slightly more complicated scenario with four pieces of data, A, B, C, and D, which can be defined as an alphabet $\mathbb{Z}$ with four letters.
The ground truth PMF is $P=\{0.1, 0.4, 0.2, 0.3\}$.
Consider two processes that combine these into two classes AB and CD.
These typify clustering algorithms, downsampling processes, discretization in visual mapping, and so on.
One process is considered to be \emph{correct}, which has a PMF for AB and CD as $R_\text{correct}=\{0.5, 0.5\}$, and another \emph{biased} process with $R_{\text{biased}}=\{0, 1\}$.
Let CG, CU, and CH be three users at the receiving end of the \emph{correct} process, and BG, BS, and BM be three other users at the receiving end of the \emph{biased} process.
The users with different knowledge exhibit different abilities to reconstruct the original scenario featuring A, B, C, D from aggregated information about AB and CD.
Similar to the \emph{good}-\emph{bad} scenario, such abilities can be captured by a PMF $Q$. For example, we have:
\begin{itemize}
  % \vspace{-2mm}
  \item CG makes random guess, $Q=\{0.25, 0.25, 0.25, 0.25\}$.
  % \vspace{-2mm}
  \item CU has useful knowledge, $Q=\{0.1, 0.4, 0.1, 0.4\}$.
  % \vspace{-2mm}
  \item CB is highly biased, $Q=\{0.4, 0.1, 0.4, 0.1\}$.
  % \vspace{-2mm}
  \item BG makes guess based on $R_{\text{biased}}$, $Q=\{0.0, 0.0, 0.5, 0.5\}$.
  % \vspace{-2mm}
  \item BS makes a small adjustment, $Q=\{0.1, 0.1, 0.4, 0.4\}$.
  % \vspace{-2mm}
  \item BM makes a major adjustment, $Q=\{0.2, 0.2, 0.3, 0.3\}$.
\end{itemize}
Figure \ref{fig:Compare-ABCD} compares the divergence values returned by the candidate measures for these six users, while the transformations from $P$ to $R_\text{correct}$ or $R_\text{biased}$, and then to $Q$ are illustrated on the right.
We can observe that $\Dncm$ and $\DnewB$ return values $<0.1$, which seem to be less intuitive. Meanwhile $\DJS$ shows a large portion of divergence from the AB category, while $\DncmA$ and $\DncmB$ show more divergence in the BC category.
In particular, for user BG, $\DncmA$ and $\DncmB$ do not show any divergence in relation to A and B, though BG clearly has reasoned A and B rather incorrectly.
$\DnewA$ and $\DnewB$ show a relatively balanced account of divergence associated with A, B, C, and D.
On balance, we give scores 5, 4, 3, 2, 1 to $\DnewA$, $\DJS$, $\DnewB$, $\DncmA$, and $\DncmB$ respectively.

% With the major shortcomings of $\Dncm (k=1, k=2)$ in this scenario, we can now focus on three commutative measures $\DJS$ and $\Dnew (k=1, k=2)$ in conjunction with two case studies.

% A major disadvantage of Eq.\,\ref{eq:CondSE} is that it requires an additional probability distribution $R$.
% In the context of this work, one would have to estimate three distributions instead of two.

% ====================
\section{Experimental Case Studies}
\label{sec:CaseStudies}
To complement the synthetic case studies in Section \ref{sec:Synthetic}, we conducted two surveys to collect some realistic examples that feature the use of knowledge in visualization.
In addition to providing instances of criteria 8 and 9 for selecting a bounded measure, the surveys were also designed to demonstrate that one could use a few simple questions to estimate the cost-benefit of visualization in relation to individual users.
% Built on the visual analysis in the previous section, we focus on three divergence measures, namely the JS divergence $\DJS$ and two versions of the new divergence, i.e., $\Dnew$ with $k=1$ and $k=2$.
% We denote $\Dnew (k=1)$ as $\mathcal{D}_1$, and $\Dnew (k=2)$ as $\mathcal{D}_2$.

% --------------------------------------------------
\subsection{Volume Visualization (Criterion 8)}
\label{sec:VolVis}
This survey, which involved ten surveyees, was designed to collect some real-world data that reflects the use of knowledge in viewing volume visualization images.
The full set of questions were presented to surveyees in the form of slides, which are included in the supplementary materials.
The full set of survey results is given in Appendix C.
The featured volume datasets were from ``The Volume Library'' \cite{Roettger:2019:web}, and visualization images were either rendered by the authors or from one of the four publications \cite{Nagy:2002:VMV,Correa:2006:TVCG,Wu:2007:TVCG,Jung:2008:web}.

The transformation from a volumetric dataset to a volume-rendered image typically features a noticeable amount of alphabet compression.
Some major algorithmic functions in volume visualization, e.g., iso-surfacing, transfer function, and rendering integral, all facilitate alphabet compression, hence information loss.

\begin{figure}[t]
\centering
\includegraphics[width=\linewidth]{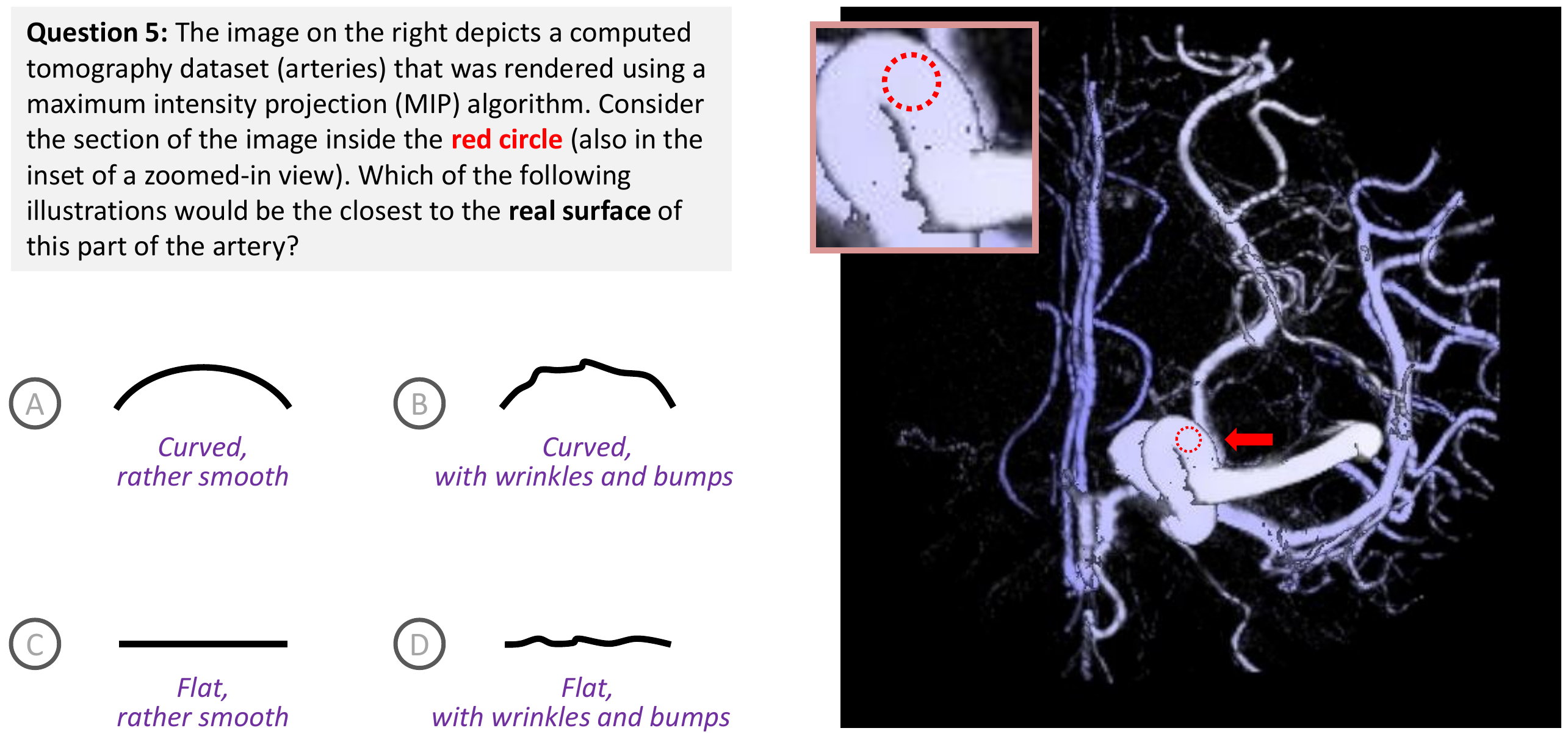}
\caption{A volume dataset was rendered using the MIP method. A question about a ``flat area'' in the image can be used to tease out a viewer's knowledge that is useful in a visualization process.}
\label{fig:Arteries}
\vspace{-4mm}
\end{figure}

In terms of rendering integral, maximum intensity projection (MIP) incurs a huge amount of information loss in comparison with the commonly-used emission-and-absorption integral \cite{Max:2010:book}.
As shown in Figure \ref{fig:Arteries}, the surface of arteries are depicted more or less in the same color.
The accompanying question intends to tease out two pieces of knowledge, ``curved surface'' and ``with wrinkles and bumps''.
Among the ten surveyees, one selected the correct answer B, eight selected the relatively plausible answer A, and one selected the doubtful answer D.

Let alphabet $\mathbb{Z}=\{\mathrm{A, B, C, D}\}$ contain the four optional answers.
One may assume a ground truth PMF $Q=\{0.1, 0.878, 0.002, 0.02\}$ since there might still be a small probability for a section of artery to be flat or smooth.
The rendered image depicts a misleading impression, implying that answer C is correct or a false PMF $F=\{0, 0, 1, 0\}$.
The amount of alphabet compression is thus $\mathcal{H}(Q) - \mathcal{H}(F) = 0.225$.

When a surveyee gives an answer to the question, it can also be considered as a PMF $P$.
With PMF $Q=\{0.1, 0.878, 0.002, 0.02\}$, different answers thus lead to different values of divergence as follows:

\begin{table}[h!]
  \vspace{-2mm}
  \centering
  \begin{tabular}{@{}l@{\hspace{5mm}}r@{\hspace{3mm}}r@{\hspace{3mm}}r@{\hspace{3mm}}r@{\hspace{3mm}}r@{}}
    \textbf{Divergence for:} & $\DJS$\; & $\DnewA$\; & $\DnewB$\; & $\DncmA$\; & $\DncmB$\;\\[1mm]
    A $(P_a = \{1, 0, 0, 0\}, Q)$: & 0.758 & 0.9087 & 0.833 & 0.926 & 0.856\\
    B $(P_b = \{0, 1, 0, 0\}, Q)$: & 0.064 & 0.1631 & 0.021 & 0.166 & 0.021\\
    C $(P_c = \{0, 0, 1, 0\}, Q)$: & 0.990 & 0.9066 & 0.985 & 0.999 & 0.997\\
    D $(P_d = \{0, 0, 0, 1\}, Q)$: & 0.929 & 0.9086 & 0.858 & 0.986 & 0.971
  \end{tabular}
  \vspace{-2mm}
%  \caption{Caption}
%  \label{tab:my_label}
\end{table}

Without any knowledge, a surveyee would select answer C, leading to the highest value of divergence in terms of any of the three measures.
Based PMF $Q$, we expect to have divergence values in the order of C $>$ D $>$ A $\gg$ B.
$\DJS$, $\DnewB$, $\DncmA$, and $\DncmB$ have produced values in that order, while $\DnewA$ indicates an order of A $>$ D $>$ C $\gg$ B.
This order cannot be interpreted easily, indicating a weakness of $\DnewA$.
% We thus score $\Dnew (k=1)$ 1 in Table \ref{tab:MultiCriteria}.

Together with the alphabet compression $\mathcal{H}(Q) - \mathcal{H}(F) = 0.225$ and the $\mathcal{H}_\text{max}$ of 2 bits, we can also calculate the informative benefit using Eq.\,\ref{eq:CBM-3}.
For surveyees with different answers, the lossy depiction of the surface of arteries brought about different amounts of benefit:

\begin{table}[h!]
  \vspace{-2mm}
  \centering
  \begin{tabular}{@{}l@{\hspace{3mm}}r@{\hspace{1.5mm}}r@{\hspace{1.5mm}}r@{\hspace{1.5mm}}r@{\hspace{1.5mm}}r@{}}
    \textbf{Benefit for:} & $\DJS$\; & $\DnewA$\; & $\DnewB$\; & $\DncmA$\; & $\DncmB$\;\\[1mm]
    A $(P_a = \{1, 0, 0, 0\}, Q)$:   & $-0.889$ & $-1.190$ & $-1.038$ & $-1.224$ & $-1.084$\\
    B $(P_b = \{0, 1, 0, 0\}, Q)$: &  $0.500$ &  $0.302$ &  $0.586$ &  $0.296$ &  $0.585$\\
    C $(P_c = \{0, 0, 1, 0\}, Q)$: & $-1.351$ & $-1.185$ & $-1.097$ & $-1.369$ & $-1.366$\\
    D $(P_d = \{0, 0, 0, 1\}, Q)$:  & $-1.230$ & $-1.189$ & $-1.088$ & $-1.343$ & $-1.314$
  \end{tabular}
  \vspace{-2mm}
%  \caption{Caption}
%  \label{tab:my_label}
\end{table}

All five sets of values indicate that only those surveyees who gave answer C would benefit from such lossy depiction produced by MIP, signifying the importance of user knowledge in visualization. However, the values returned for A, C, and D by $\DnewA$ are almost indistinguishable and in an undesirable order.

One may also consider the scenarios where flat or smooth surfaces are more probable.
For example, if the ground truth PMF were $Q'=\{0.30, 0.57, 0.03, 0.10\}$ and $\mathcal{H}(Q') = 1.467$, the amounts of benefit would become:

\begin{table}[h!]
  \vspace{-2mm}
  \centering
  \begin{tabular}{@{}l@{\hspace{3mm}}r@{\hspace{1.5mm}}r@{\hspace{1.5mm}}r@{\hspace{1.5mm}}r@{\hspace{1.5mm}}r@{}}
    \textbf{Benefit for:} & $\DJS$\; & $\DnewA$\; & $\DnewB$\; & $\DncmA$\; & $\DncmB$\;\\[1mm]
    A $(P_a = \{1, 0, 0, 0\}, Q')$: &  $0.480$ &  $0.086$ &  $0.487$ & $-0.064$ &  $0.317$\\
    B $(P_b = \{0, 1, 0, 0\}, Q')$: &  $0.951$ &  $0.529$ &  $1.044$ &  $0.435$ &  $0.978$\\
    C $(P_c = \{0, 0, 1, 0\}, Q')$: & $-0.337$ & $-0.038$ &  $0.212$ & $-0.489$ & $-0.446$\\
    D $(P_d = \{0, 0, 0, 1\}, Q')$: & $-0.049$ & $-0.037$ &  $0.257$ & $-0.385$ & $-0.245$
  \end{tabular}
  \vspace{-2mm}
%  \caption{Caption}
%  \label{tab:my_label}
\end{table}

\noindent Because the ground truth PMF $Q'$ would be less certain, the knowledge of ``curved surface'' and ``with wrinkles and bumps'' would become more useful.
Further, because the probability of flat and smooth surfaces would have increased, an answer C would not be as bad as when it is with the original PMF $Q$.
Among the five measures, $\DJS$, $\DnewB$, $\DncmA$, and $\DncmB$ returned values indicating the same order of benefit, i.e., B $>$ A $>$ D $>$ C, which is consistent with PMF $Q'$.
Only $\DnewA$ orders C and D differently.

We can also observe that these measures occupy different ranges of real values.
$\DnewB$ appears to be more generous in valuing the benefit of visualization, while $\DncmA$ is less generous. We will examine this phenomenon with a more compelling example in Section \ref{sec:London}. 

% The above example of MIP rendering shows that to those users with the appropriate knowledge, the missing information in a visualization image is not really ``lost''.
% Using the categorization of visual multiplexing \cite{Chen:2014:CGF}, the information about ``curved surface'' and ``with wrinkles and bumps'' is conveyed using a \emph{hollow visual channel}.
% Volume visualization features some other forms of visual multiplexing.
% The viewers' ability to de-multiplex depends on their knowledge, which can now be estimated quantitatively.

\begin{figure}[t]
\centering
\includegraphics[width=\linewidth]{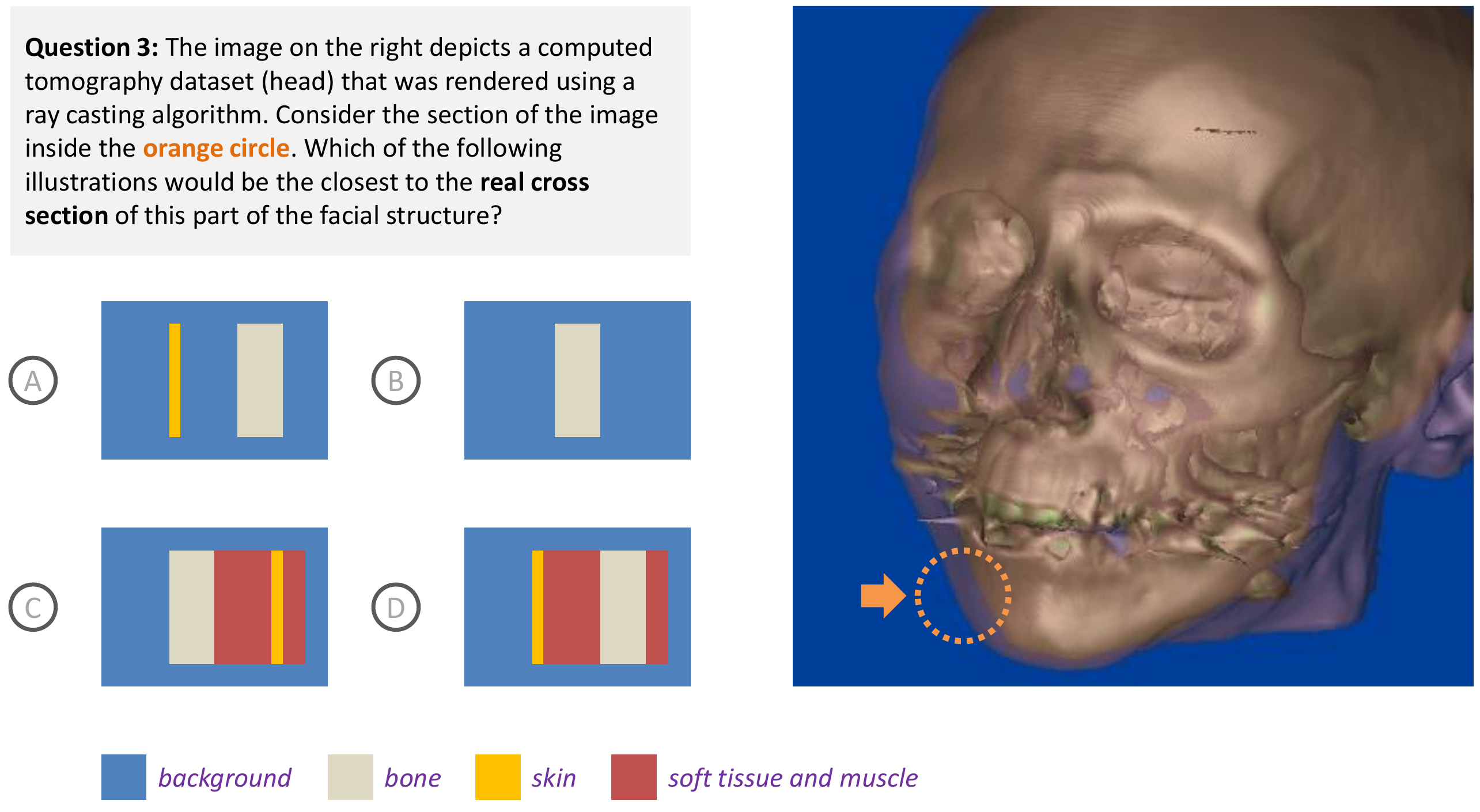}
\caption{Two iso-surfaces of a volume dataset were rendered using the ray casting method. A question about the tissue configuration in the orange circle can tease out a viewer's knowledge about the translucent depiction and the missing information.}
\label{fig:Isosurfaces}
\vspace{-4mm}
\end{figure}

Figure \ref{fig:Isosurfaces} shows another volume-rendered image used in the survey.
Two iso-surfaces of a head dataset are depicted with \emph{translucent occlusion}, which is a type of visual multiplexing \cite{Chen:2014:CGF}.
Meanwhile, the voxels for soft tissue and muscle are not depicted at all, which can also been regarded as using a \emph{hollow visual channel}. 
The visual representation has been widely used, and the viewers are expected to use their knowledge to infer the 3D relationships between the two iso-surfaces as well as the missing information about soft tissue and muscle.
The question that accompanies the figure is for estimating such knowledge.

Although the survey offers only four options, it could in fact offer many other configurations as optional answers.
Let us consider four color-coded segments similar to the configurations in answers C and D.
Each segment could be one of four types: bone, skin, soft tissue and muscle, or background.
There are a total of $4^4=256$ configurations.
If one had to consider the variation of segment thickness, there would be many more options.
Because it would not be appropriate to ask a surveyee to select an answer from 256 options, a typical assumption is that the selected four options are representative.
In other words, considering that the 256 options are letters of an alphabet, any unselected letter has a probability similar to one of the four selected options. 

For example, we can estimate a ground truth PMF $Q$ such that among the 256 letters,
\begin{itemize}
  % \vspace{-2mm}
  \item Answer A and four other letters have a probability 0.01,
  % \vspace{-2mm}
  \item Answer B and 64 other letters have a probability 0.0002,
  % \vspace{-2mm}
  \item Answer C and 184 other letters have a probability 0.0001,
  % \vspace{-2mm}
  \item Answer D has a probability 0.9185.
\end{itemize}
We have the entropy of this alphabet $\mathcal{H}(Q)=0.85$.
Similar to the previous example, we can estimate the values of divergence as:

\begin{table}[h!]
  \vspace{-2mm}
  \centering
  \begin{tabular}{@{}l@{\hspace{3mm}}r@{\hspace{2.3mm}}r@{\hspace{2.3mm}}r@{\hspace{2.3mm}}r@{\hspace{2.3mm}}r@{}}
    Divergence for: & $\DJS$\; & $\DnewA$\; & $\DnewB$\; & $\DncmA$\; & $\DncmB$\;\\[1mm]
    A: $P = \{1, \ddddot{_4}, 0, \ddddot{_{64}}, 0, \ddddot{_{184}}, 0\}$
        & 0.960 & 0.933 & 0.903 & 0.993 & 0.986\\
    B: $P = \{0, \ddddot{_4}, 1, \ddddot{_{64}}, 0, \ddddot{_{184}}, 0\}$
        & 0.999 & 0.932 & 0.905 & 1.000 & 1.000\\
    C: $P = \{0, \ddddot{_4}, 0, \ddddot{_{64}}, 1, \ddddot{_{184}}, 0\}$
        & 0.999 & 0.932 & 0.905 & 1.000 & 1.000\\
    D: $P = \{0, \ddddot{_4}, 0, \ddddot{_{64}}, 0, \ddddot{_{184}}, 1\}$
        & 0.042 & 0.109 & 0.009 & 0.113 & 0.010
  \end{tabular}
  \vspace{-2mm}
%  \caption{Caption}
%  \label{tab:my_label}
\end{table}

\noindent where $\ddddot{_{n}}$ denotes $n$ zeros.
$\DJS$, $\DnewB$, $\DncmA$ and $\DncmB$ returned values indicating the same order of divergence, i.e., C $\sim$ B $>$ A $\gg$ D, which is consistent with PMF $Q'$.
Only $\DnewA$ returns an order A $>$ B $\sim$ C $\gg$ D.
This reinforces the observation by the previous example (i.e., Figure \ref{fig:Arteries}) about the characteristics of ordering of the five measures.

For both examples (Figs. \ref{fig:Arteries} and \ref{fig:Isosurfaces}), because both $\DJS$, $\DnewB$, $\DncmA$ and $\DncmB$ have consistently returned sensible values, we give a score of 5 to each of them in Table \ref{tab:MultiCriteria}.
$\DnewA$ appears to be often incompatible with other measures in terms of ordering, we score $\DnewA$ 1. 

% With the maximum entropy being 8 bits, we can estimate the amounts of informative benefit as:
%
% \begin{align*}
%  &\text{with } \mathcal{D}_{\text{JS}},
%    &\text{A}: -6.826, \text{ B}: -7.139, \text{ C}: -7.144, \text{ D}: 0.514\\
%  &\text{with } \DnewB,
%    &\text{A}: -6.374, \text{ B}: -6.392, \text{ C}: -6.392, \text{ D}: 0.777
% \end{align*}
%

% such as \emph{partial occlusion}, , \emph{continuous field}, \emph{shifting visual channels}, and \emph{assuming a priori knowledge}

\begin{figure*}[t]
  \centering
  \includegraphics[height=46mm]{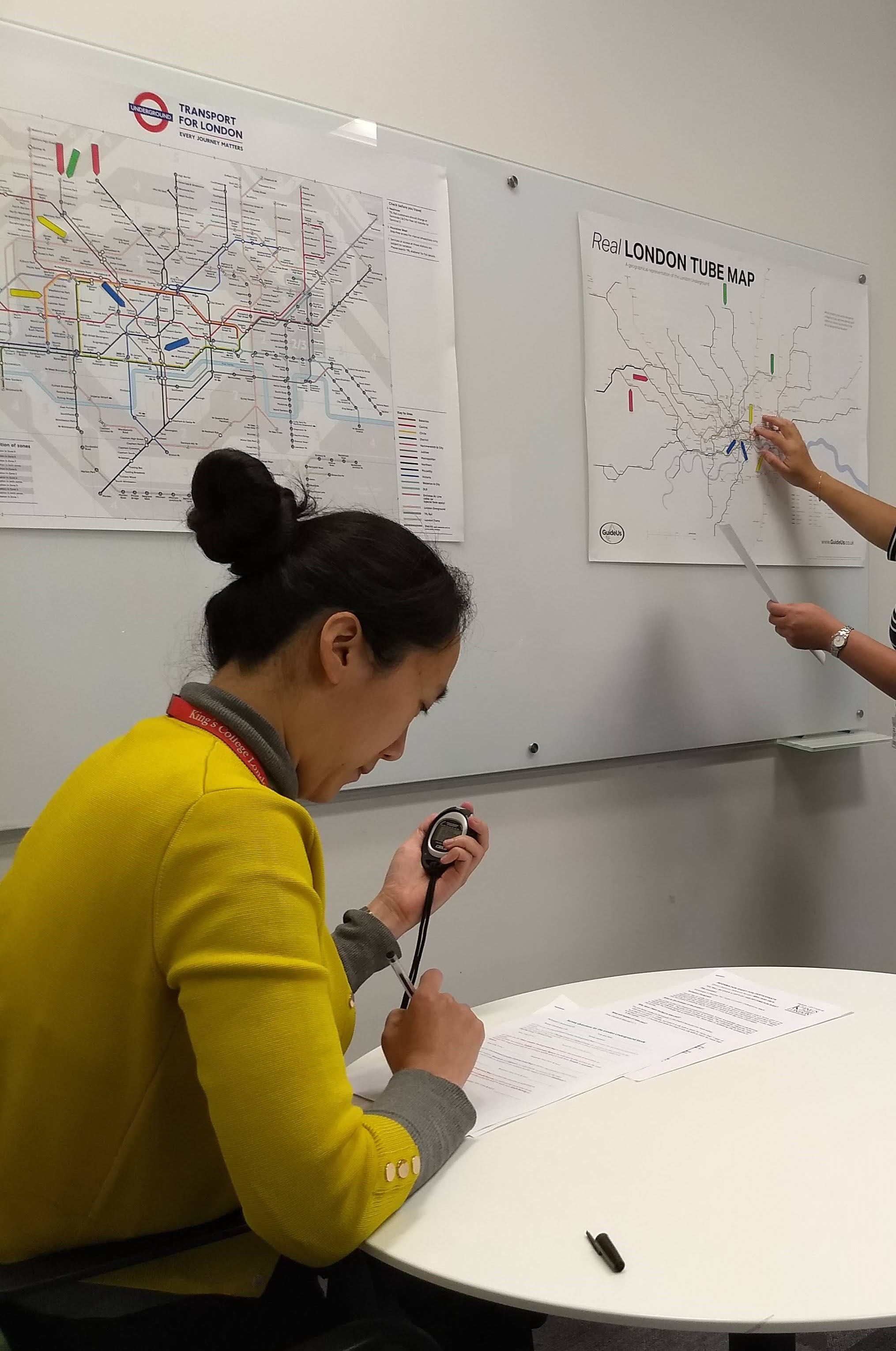}
  \includegraphics[height=46mm]{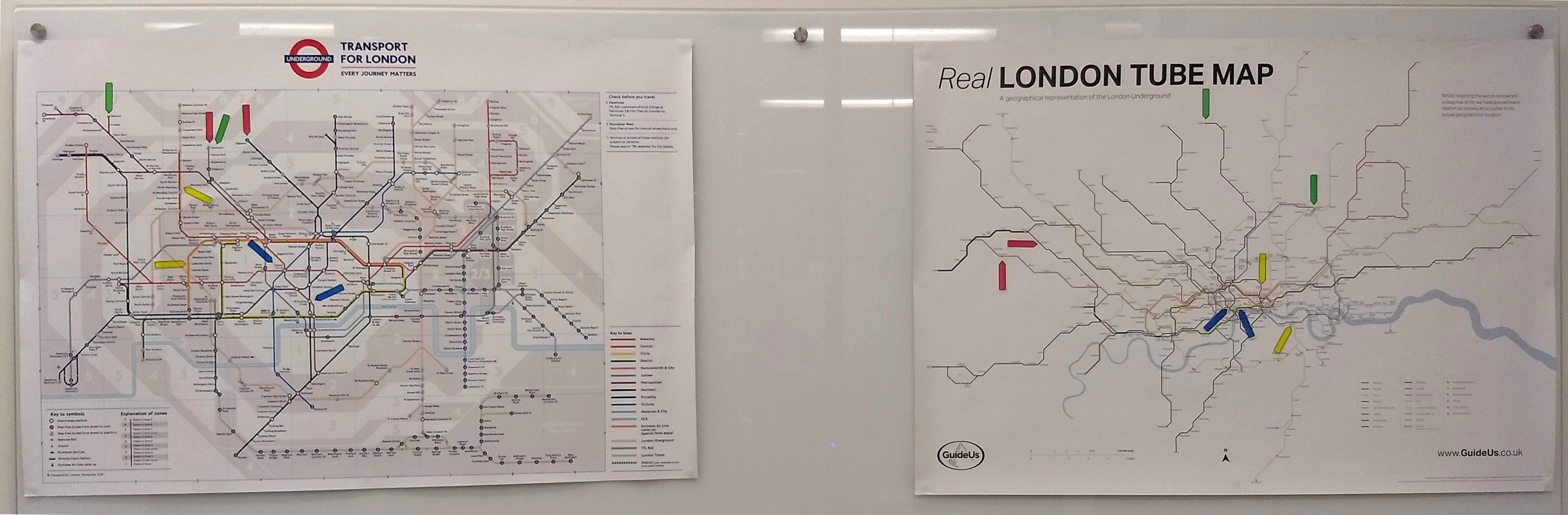}
  \caption{A survey for collecting data that reflects the use of some knowledge in viewing two types of London underground maps.}
  \label{fig:LondonMaps}
  \vspace{-4mm}
\end{figure*}

% --------------------------------------------------
\subsection{London Underground Map (Criterion 9)}
\label{sec:London}
This survey was designed to collect some real-world data that reflects the use of knowledge in viewing different London underground maps.
It involved sixteen surveyees, twelve at King's College London (KCL) and four at University of Oxford.
Surveyees were interviewed individually in a setup as shown in Figure \ref{fig:LondonMaps}.  
Each surveyee was asked to answer 12 questions using either a geographically-faithful map or a deformed map, followed by two further questions about their familiarity of a metro system and London.
A \pounds5 Amazon voucher was offered to each surveyee as an appreciation of their effort and time.
The survey sheets and the full set of survey results are given in Appendix D.

Harry Beck first introduced geographically-deformed design of the London underground maps in 1931.
Today almost all metro maps around the world adopt this design concept.
Information-theoretically, the transformation of a geographically-faithful map to such a geographically-deformed map causes a significant loss of information.
Naturally, this affects some tasks more than others.

For example, the distances between stations on a deformed map are not as useful as in a faithful map.
The first four questions in the survey asked surveyees to estimate how long it would take to walk (i) from \emph{Charing Cross} to \emph{Oxford Circus},
(ii) from \emph{Temple} and \emph{Leicester Square},
(iii) from \emph{Stanmore} to \emph{Edgware}, and
(iv) from \emph{South Rulslip} to \emph{South Harrow}.
On the deformed map, the distances between the four pairs of the stations are all about 50mm.
On the faithful map, the distances are (i) 21mm, (ii) 14mm, (iii) 31mm, and (iv) 53mm respectively. 
According to the Google map, the estimated walk distance and time are (i) 0.9 miles, 20 minutes; (ii) 0.8 miles, 17 minutes; (iii) 1.6 miles, 32 minutes; and (iv) 2.2 miles, 45 minutes respectively.

The average range of the estimations about the walk time by the 12 surveyees at KCL are: (i) 19.25 [8, 30], (ii) 19.67 [5, 30], (iii) 46.25 [10, 240], and (iv) 59.17 [20, 120] minutes.
The estimations by the four surveyees at Oxford are:
(i) 16.25 [15, 20], (ii) 10 [5, 15], (iii) 37.25 [25, 60], and (iv) 33.75 [20, 60] minutes.
The values correlate better to the Google estimations than what would be implied by the similar distances on the deformed map.
Clearly some surveyees were using some knowledge to make better inference.

Let $\mathbb{Z}$ be an alphabet of integers between 1 and 256.
The range is chosen partly to cover the range of the answers in the survey, and partly to round  
up the maximum entropy $\mathbb{Z}$ to 8 bits.
For each pair of stations, we can define a PMF using a skew normal distribution peaked at the Google estimation $\xi$.
As an illustration, we coarsely approximate the PMF as $Q = \{q_i \; | \; 1 \le i \le 256 \}$, where
\begin{equation*}
  q_i = \begin{cases}
    0.01/236    & \text{if }\; 1 \le i \le \xi - 8  \hspace{10mm} (\textit{wild guess})\\
    0.026        & \text{if }\; \xi - 7 \le i \le \xi - 3 \quad \;\;\, (\textit{close})\\
    0.12        & \text{if }\; \xi - 2 \le i \le \xi + 2 \quad \;\;\, (\textit{spot on})\\
    0.026        & \text{if }\; \xi + 3 \le i \le \xi + 12 \quad (\textit{close})\\
    0.01/236    & \text{if }\; \xi + 13 \le i \le 256  \quad \;\;\, (\textit{wild guess})
  \end{cases}
\end{equation*}
Using the same way in the previous case study, we can estimate the divergence and the benefit of visualization for an answer in each range. Recall our observation of the phenomenon in Section \ref{sec:VolVis} that the measurements by $DJS$, $\DnewA$, $\DnewB$, $\DncmA$ and $\DncmB$ occupy different ranges of values, with $\DnewB$ be the most generous in measuring the benefit of visualization.
With the entropy of the alphabet as $\mathcal{H}(Q) \approx 3.6$ bits and the maximum entropy being 8 bits, the benefit values obtained for this example exhibit a similar but more compelling pattern:%

\begin{table}[h!]
  \vspace{-2mm}
  \centering
  \begin{tabular}{@{}lrrrrr@{}}
    Benefit for: & $\DJS$\; & $\DnewA$\; & $\DnewB$\; & $\DncmA$\; & $\DncmB$\;\\[1mm]
    \emph{spot on}    & $-1.765$ & $-0.418$ & \textbf{0.287} & $-3.252$ & $-2.585$\\
    \emph{close}      & $-3.266$ & $-0.439$ & \textbf{0.033} & $-3.815$ & $-3.666$\\
    \emph{wild guess} & $-3.963$ & $-0.416$ & $-0.017$ & $-3.966$ & $-3.965$
  \end{tabular}
  \vspace{-2mm}
%  \caption{Caption}
%  \label{tab:my_label}
\end{table}

\noindent Only $\DnewB$ has returned positive benefit values for \emph{spot on} and \emph{close} answers. Since it is not intuitive to say that those surveyees who gave good answers benefited from visualization negatively, clearly only the measurements returned by $\DnewB$ are intuitive. In addition, the ordering resulting from $\DnewA$ is again inconsistent with others.

For instance, surveyee P9, who has lived in a city with a metro system for a period of 1-5 years and lived in London for several months, made similarly good estimations about the walking time with both types of underground maps.
With one \emph{spot on} answer and one \emph{close} answer under each condition, the estimated benefit on average is $0.160$ bits if one uses $\DnewB$ and is negative if one uses any of the other four measures.
Meanwhile, surveyee P3, who has lived in a city with a metro system for two months, provided all four answers in the \emph{wild guess} category, leading to negative benefit values by all five measures.

Similarly, among the first set of four questions in the survey, Questions 1 and 2 are about stations near KCL, and Questions 3 and 4 are about stations more than 10 miles away from KCL.
The local knowledge of the surveyees from KCL clearly helped their answers.
Among the answers given by the twelve surveyees from KCL,
\begin{itemize}
    % \vspace{-2mm}
    \item For Question 1, four \emph{spot on}, five \emph{close}, and three \emph{wild guess} --- the average benefit is $-2.940$ with $\mathcal{D}_{\text{JS}}$ or $0.105$ with $\DnewB$.
    % \vspace{-2mm}
    \item For Question 2, two \emph{spot on}, nine \emph{close}, and one \emph{wild guess} ---
    the average benefit is $-3.074$ with $\mathcal{D}_{\text{JS}}$ or $0.071$ with $\DnewB$.
    % \vspace{-2mm}
    \item For Question 3, three \emph{close}, and nine \emph{wild guess} ---
    the average benefit is $-3.789$ with $\mathcal{D}_{\text{JS}}$ or $-0.005$ with $\DnewB$.
    % \vspace{-2mm}
    \item For Question 4, two \emph{spot on}, one \emph{close}, and nine \emph{wild guess} --- the average benefit is with $-3.539$ $\mathcal{D}_{\text{JS}}$ or $0.038$ with $\DnewB$.
\end{itemize}

The average benefit values returned by $\DnewA$, $\DncmA$, and $\DncmB$ are all negative for these four questions. Hence, unless $\DnewB$ is used, all other measures would semantically imply that both types of the London underground maps would have negative benefit. 
We therefore give $\DnewB$ a 5 score and $\DJS$, $\DncmA$, and $\DncmB$ a 3 score each in Table \ref{tab:MultiCriteria}.
We score $\DnewA$ 1 as it also exhibits an ordering issue.  

When we consider answering each of Questions 1$\sim$4 as performing a visualization task, we can estimate the cost-benefit ratio of each process.
As the survey also collected the time used by each surveyee in answering each question, the cost in Eq.\,\ref{eq:CBM-1} can be approximated with the mean response time.
For Questions 1$\sim$4, the mean response times by the surveyees at KCL are 9.27, 9.48, 14.65, and 11.40 seconds respectively.
Using the benefit values based on $\DnewB$, the cost-benefit ratios are thus 0.0113, 0.0075, -0.0003, and 0.0033 bits/second respectively. While these values indicate the benefits of the local knowledge used in answering Questions 1 and 2, they also indicate that when the local knowledge is absent in the case of Questions 3 and 4, the deformed map (i.e., Question 3) is less cost-beneficial.

% ====================
\section{Conclusions}
\label{sec:Conclusions}
This paper is a follow-on paper that continues an investigation into the need to improve the mathematical formulation of an information-theoretic measure for analyzing the cost-benefit of visualization as well as other processes in a data intelligence workflow \cite{Chen:2016:TVCG}.
The concern about the original measure is its unbounded term based on the KL-divergence.
The preceding paper (in the supplementary materials) studied eight candidate measures and use conceptual evaluation to narrow the options down to five, providing important evidence to the multi-criteria analysis of thees candidate measures.

% In this work, we used two synthetic and two experimental case studies to obtain some data to allow us observe the behaviours of the candidate measures.
% One consists of questions about volume visualizations, while the other features visualization tasks performed in conjunction with two types of London Underground maps.
% The case studies allowed us to test some most promising candidate measures with the real world data collected in the two surveys, providing important evidence to two important aspects of the multi-criteria analysis.
%
From Table \ref{tab:MultiCriteria}, we can observe the process of narrowing down from eight candidate measures to five measures, and then to one.
Taking the importance of the criteria into account, we consider that candidate $\Dnew (k=2)$ is ahead of $\DJS$, critically because $\DJS$ often yields negative benefit values even when the benefit of visualization is clearly there.
We therefore propose to revise the original cost-benefit ratio in \cite{Chen:2016:TVCG} to the following:
\begin{equation} \label{eq:CBM-4}
  \begin{split}
          \frac{\text{Benefit}}{\text{Cost}} &= \frac{\text{Alphabet Compression} - \text{Potential Distortion}}{\text{Cost}}\\
          &= \frac{\SE(\mathbb{Z}_i) - \SE(\mathbb{Z}_{i+1}) - \SE_{\text{max}}(\mathbb{Z}_i) \mathcal{D}^2_{\text{new}}(\mathbb{Z}'_i || \mathbb{Z}_i)}{\text{Cost}}
  \end{split}
\end{equation}

This cost-benefit measure was developed in the field of visualization, for optimizing visualization processes and visual analytics workflows.
Its broad interpretation may include data intelligence workflows in other contexts \cite{Chen:2020:OUP}.
The measure has now been improved by using visual analysis and with the survey data collected in the context of visualization applications.

The history of measurement science \cite{Klein:2012:book} informs us that proposals for metrics, measures, and scales will continue to emerge in visualization, typically following the arrival of new theoretical understanding, new observational data, new measurement technology, and so on. As measurement is one of the driving forces in science and technology. We shall welcome such new measurement development in visualization.

The work presented in this paper and the preceding paper does not indicate a closed chapter, but an early effort to be improved frequently in the future.
For example, future work may discover measures that have better mathematical properties than $\Dnew$ and $\DJS$, or future experimental observation may evidence that $\DJS$ offer more intuitive explanations than $\Dnew$ in other case studies.
In particular, we would like to continue our theoretical investigation into the mathematical properties of $\Dnew$.

``Measurement is not an end but a means in the process of description, differentiation, explanation, prediction, diagnosis, decision making, and the like'' \cite{Pedhazur:1991:book}.
Having a bounded cost-benefit measure offers many new opportunities of developing tools for aiding the measurement and using such tools in practical applications, especially in visualization and visual analytics.

% The authors wish to thank A, B, and C. This work was supported
% in part by a grant from XYZ (\# 12345-67890).}

%\bibliographystyle{abbrv}
% \bibliographystyle{abbrv-doi}
% \bibliographystyle{abbrv-doi-narrow}
% \bibliographystyle{abbrv-doi-hyperref}
% \bibliographystyle{abbrv-doi-hyperref-narrow}

\bibliographystyle{eg-alpha-doi}  
\bibliography{EstimatePD}
% biblatex with biber
% \printbibliography 

\clearpage
\newpage
\noindent\huge%
\textsc{\textsf{\textbf{Appendices}}}

\noindent\LARGE%
\textbf{A Bounded Measure for Estimating\\the Benefit of Visualization}

\noindent\large%
~\\
Min Chen, University of Oxford, UK\\
Alfie Abdul-Rahman, King's College London, UK\\
Deborah Silver, Rutgers University, USA\\
Mateu Sbert, University of Girona, Spain\\

\normalsize%

\appendix
% =================================================
\section{\textbf{Explanation of the Original Cost-Benefit Measure}}
\label{app:OriginalTheory}
This appendix contains an extraction from a previous publication \cite{Chen:2019:CGF}, which provides a relatively concise but informative description of the cost-benefit ratio proposed in \cite{Chen:2016:TVCG}. The inclusion of this is to minimize the readers' effort to locate such an explanation. The extraction has been slightly modified.
In addition, at the end of this appendix, we provide a relatively informal and somehow conversational discussion about using this measure to explain why visualization is useful.

Chen and Golan introduced an information-theoretic metric for measuring the cost-benefit ratio of a visual analytics (VA) workflow or any of its component processes \cite{Chen:2016:TVCG}.
The metric consists of three fundamental measures that are abstract representations of a variety of qualitative and quantitative criteria used in practice, including
operational requirements (e.g., accuracy, speed, errors, uncertainty, provenance, automation),
analytical capability (e.g., filtering, clustering, classification, summarization),
cognitive capabilities (e.g., memorization, learning, context-awareness, confidence), and so on.
The abstraction results in a metric with the desirable mathematical simplicity \cite{Chen:2016:TVCG}.
The qualitative form of the metric is as follows:
\begin{equation}
\label{eq:CBR}
\frac{\textit{Benefit}}{\textit{Cost}} = \frac{\textit{Alphabet Compression} - \textit{Potential Distortion}}{\textit{Cost}}
\end{equation}

The metric describes the trade-off among the three measures:

\begin{itemize}
\vspace{-1mm}
\item
\emph{Alphabet Compression} (AC) measures the amount of entropy reduction (or information loss) achieved by a process.
As it was noticed in \cite{Chen:2016:TVCG}, most visual analytics processes (e.g., statistical aggregation, sorting, clustering, visual mapping, and interaction), feature many-to-one mappings from input to output, hence losing information.
Although information loss is commonly regarded harmful, it cannot be all bad if it is a general trend of VA workflows.
Thus the cost-benefit metric makes AC a positive component.
\vspace{-1mm}
\item
\emph{Potential Distortion} (PD) balances the positive nature of AC by measuring the errors typically due to information loss. Instead of measuring mapping errors using some third party metrics, PD measures the potential distortion when one reconstructs inputs from outputs.
The measurement takes into account humans' knowledge that can be used to improve the reconstruction processes. For example, given an average mark of 62\%, the teacher who taught the class can normally guess the distribution of the marks among the students better than an arbitrary person.
\vspace{-1mm}
\item
\emph{Cost} (Ct) of the forward transformation from input to output and the inverse transformation of reconstruction provides a further balancing factor in the cost-benefit metric in addition to the trade-off between AC and PD. In practice, one may measure the cost using \emph{time} or a monetary measurement.
\end{itemize}

\vspace{2mm}
\noindent\textbf{Why is visualization useful?}
There have been many arguments about why visualization is useful. 
Streeb et al. collected a large number of arguments and found many arguments are in conflict with each other \cite{Streeb:2019:TVCG}.
Chen and Edwards presented an overview of schools of thought in the field of visualization, and showed that the ``why'' question is a bone of major contention \cite{Chen:2020:book}.

The most common argument about ``why'' question is because visualization offers insight or helps humans to gain insight. When this argument is used outside the visualization community, there are often counter-arguments that statistics and algorithms can offer insight automatically and often with better accuracy and efficiency. There are also concerns that visualization may mislead viewers, which cast further doubts about the usefulness of visualization, while leading to a related argument that ``visualization must be accurate'' in order for it to be useful.
The accuracy argument itself is not bullet-proof since there are many types of uncertainty in a visualization process, from uncertainty in data, to that caused by visual mapping, and to that during perception and cognition \cite{Dasgupta:2012:CGF}.
Nevertheless, it is easier to postulate that visualization must be accurate, as it seems to be counter-intuitive to condone the idea that ``visualization can be inaccurate,'' not mentioning the idea of ``visualization is normally inaccurate,'' or ``visualization should be inaccurate.''

The word ``inaccurate'' is itself an abstraction of many different types of inaccuracy.
Misrepresentation truth is a type of inaccuracy.
Such acts are mostly wrong, but some (such as wordplay and sarcasm) may cause less harm.
Converting a student's mark in the range of [0, 100] to the range of [A, B, C, D, E, F] is another type of inaccuracy.
This is a common practice, and must be useful.
From an information-theoretic perspective, these two types of inaccuracy are information loss.

In their paper \cite{Chen:2016:TVCG}, Chen and Golan observed that statistics and algorithms usually lose more information than visualization. Hence, this provides the first hint about the usefulness of visualization. They also noticed that like wordplay and sarcasm, the harm of information loss can be alleviated by knowledge. For someone who can understand a workplay (e.g., a pun) or can sense a sarcastic comment, the misrepresentation can be corrected by that person at the receiving end. This provides the second hint about the usefulness of visualization because any ``misrepresentation'' in visualization may be corrected by a viewer with appropriate knowledge.

On the other hand, statistics and algorithms are also useful, and sometimes more useful than visualization. Because statistics and algorithms usually cause more information loss, some aspects of information loss must be useful.
One important merit of losing information in one process is that the succeeding process has less information to handle, and thus incurs less cost.
This is why Chen and Golan divided information loss into two components, a positive component called alphabet compression and a negative component called potential distortion \cite{Chen:2016:TVCG}.

The positive component explains why statistics, algorithms, visualization, and interaction are useful because they all lose information.
The negative component explains why they are sometimes less useful because information loss may cause distortion during information reconstruction.
Both components are moderated by the cost of a process (i.e., statistics, algorithms, visualization, or interaction) in losing information and reconstructing the original information.
Hence, given a dataset, the best visualization is the one that loses most information while causing the least distortion.
This also explains why visual abstraction is effective when the viewers have adequate knowledge to reconstruct the lost information and may not be effective otherwise \cite{Viola:2019:book}.

The central thesis by Chen and Golan \cite{Chen:2016:TVCG} may appear to be counter-intuitive to many as it suggests ``inaccuracy is a good thing'', partly because the word ``inaccuracy'' is an abstraction of many meanings and itself features information loss. Perhaps the reason for the conventional wisdom is that it is relatively easy to think that ``visualization must be accurate''. To a very small extent, this is a bit like the easiness to think ``the earth is flat'' a few centuries ago, because the evidence for supporting that wisdom was available everywhere, right in front of everyone at that time.
Once we step outside the field of visualization, we can see the phenomena of inaccuracy everywhere, in statistics and algorithms as well as in visualization and interaction.
All these suggest that ``the earth may not be flat,'' or ``inaccuracy can be a good thing.''

In summary, the cost-benefit measure by Chen and Golan \cite{Chen:2016:TVCG} explains that when visualization is useful, it is because visualization has a better trade-off than simply reading the data, simply using statistics alone, or simply relying on algorithms alone.
The ways to achieve a better trade-off include: (i) visualization may lose some information to reduce the human cost in observing and analyzing the data, (ii) it may lose some information since the viewers have adequate knowledge to recover such information or can acquire such knowledge at a lower cost, (iii) it may preserve some information because it reduces the reconstruction distortion in the current and/or succeeding processes, and (iv) it may preserve some information because the viewers do not have adequate knowledge to reconstruct such information or it would cost too much to acquire such knowledge.

\begin{figure*}[t]
  \centering
  \includegraphics[width=178mm]{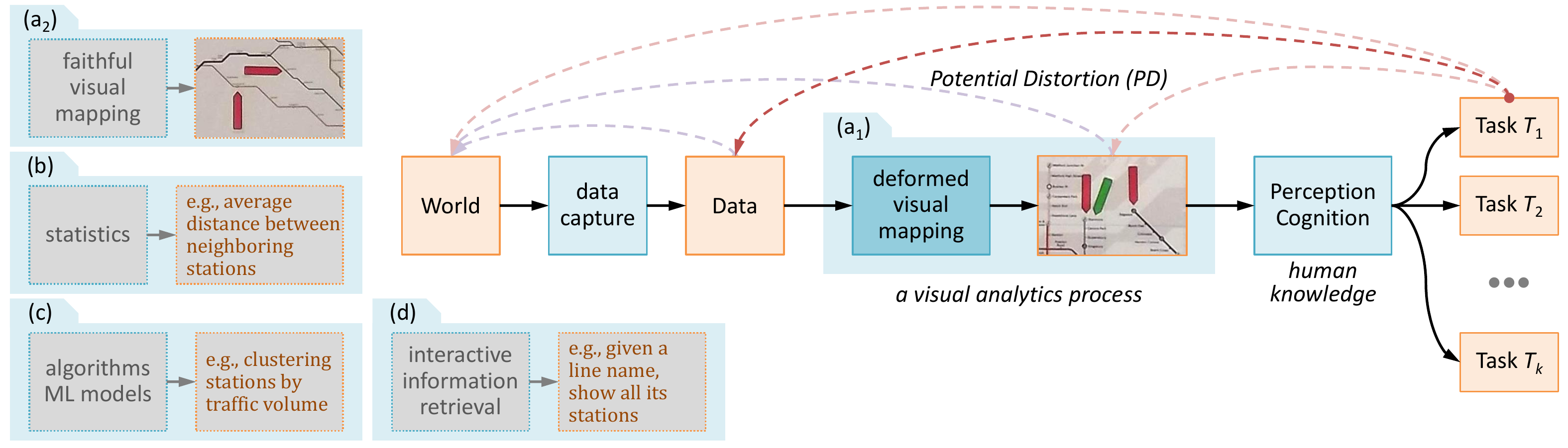}
  \caption{A visual analytics workflow features a general trend of alphabet compression from left (World) to right (Tasks). The potential distortion compares at an information space reconstructed based on the output with the original input information space. When we place different processes (i.e., (a$_1$), (a$_2$), (b), (c), and (d)), in the workflow, we can appreciate that statistics, algorithms, visualization, and interaction have different levels of alphabet compression, potential distortion, and cost.}
  \label{fig:Concept}
  \vspace{-4mm}
\end{figure*}

% ====================
\section{\textbf{How Tasks and Users are Featured in the Cost-benefit Measures?}}
\label{app:TasksUsers}
Whilst hardly anyone in the visualization community would support any practice intended to deceive viewers, there have been many visualization techniques that inherently cause distortion to the original data.
The deformed London underground map in Figure \ref{fig:InfoLoss} shows such an example.
The distortion in this example is largely caused by many-to-one mappings.
A group of lines that would be shown in different lengths in a faithful map are now shown with the same length.
Another group of lines that would be shown with different geometric shapes are now shown as the same straight line.
In terms of information theory, when the faithful map is transformed to the deformed, a good portion of information has been lost because of these many-to-one mappings.

In fact, there are many other forms of information loss. For example, when a high-resolution data variable (e.g., an integer in the range [0, 10,000]) is visually encoded as a bar in a bar chart that is restricted to a height of 1,000 pixels, about every 10 values are mapped onto the same height in terms of pixels.
It is unlikely that humans can precisely identify the height of each bar at the pixel resolution.
Likely a viewer may perceive a height of 833 pixels to be the same as one with 832 pixels or 834 pixels, which is also a many-to-one mapping.
When multivariate data records are encoded as glyphs, there is usually a significant amount of information loss. 
As we will discuss later in this paper, in volume visualization, when a sequence of $n$ voxel values are transformed to a single pixel value, as long as $n$ is a reasonably large value, a huge amount of information loss is almost guaranteed to happen.

Despite the ubiquitous phenomenon of information loss in visualization, it has been difficult for many of us to contemplate the idea that information loss may be a good thing. In particular, one theory based on an algebraic framework defines three principles that formalize the notion of graphical integrity to prevent such information loss \cite{Kindlmann:2014:TVCG}.
When one comes across an effective visualization but featuring noticeable information loss, the typical answer is that it is task-dependent, and the lost information is not useful to the task concerned.
When a visualization is evaluated, a common critique is about information loss, such as inadequate resolution, view obstruction, distorted representation, which are also characteristics of the aforementioned glyphs, volume rendering, and deformed metro map respectively.

The common phrase that ``the appropriateness of information loss depends on tasks'' is not an invalid explanation. But on its own, this explanation is not adequate, because:
\begin{itemize}
    \item The appropriateness depends on many attributes of a task, such as the selection of variables in the data and their encoded visual resolution required to complete a task satisfactorily, and the time allowed to complete a task;
    \item The appropriateness depends also on other factors in a visualization process, such as the original data resolution, the viewer's familiarity of the data, the extra information that is not in the data but the viewer knows, and the available visualization resources;
    \item The phrase creates a gray area as to whether information loss is allowed or not, and when or where one could violate some principles such as those principles in \cite{Kindlmann:2014:TVCG}. 
\end{itemize}

Partly inspired by the above puzzling dilemma in visualization, and partly by a similar conundrum in economics ``what is the most appropriate resolution of time series for an economist'', Chen and Golan proposed an information-theoretic cost-benefit ratio for measuring various factors involved in visualization processes \cite{Chen:2016:TVCG}.
Because this cost-benefit ratio can measure some abstract characteristics of ``data'', ``visualization'', ``information loss'', ``knowledge'', and ``task'' using the most fundamental information-theoretic unit \emph{bit}, it provides a means to define their relationship coherently.
Before we introduce the cost-benefit ratio mathematically in the next section, we first give its qualitative version below and provide a qualitative explanation for readers who are not familiar with information theory:
\begin{equation} \label{eq:CBM-1a}
    \frac{\text{Benefit}}{\text{Cost}} = \frac{\text{Alphabet Compression} - \text{Potential Distortion}}{\text{Cost}}
\end{equation}

Chen and Golan noticed that not only visualization processes lose information, but other data intelligence processes also lose information. For example, when statistics is used to down-sample a time series, or to compute its statistical properties, there is a substantial amount of information loss; when an algorithm groups data points into clusters or sort them according to a key variable, there is information loss; and when a computer system asks a user to confirm an action, there is information loss in the computational process \cite{Chen:2018:arXiv}. They also noticed that almost all decision tasks, the number of decision options is usually rather small. In terms of information theoretic quantities, the amount of information associated with a decision task is usually much lower than the amount of information associated with the data entering a data intelligence workflow.
They concluded that this general trend of information reduction must be a positive thing for any data intelligence workflows.
They referred to the amount of information reduction as \emph{Alphabet Compression} (AC) and made it a positive contribution to the \emph{benefit} term in Eq.\,\ref{eq:CBM-1a}.

Figure \ref{fig:Concept} shows an example of a simple visual analytics workflow, where at the moment, the visual analytics process is simply a visualization process, (a$_1$), for viewing a deformed London underground map. There can be many possible visualization tasks, such as counting the number of stops between two stations, searching for a suitable interchange station, and so on. From the workflow in Figure \ref{fig:Concept}, one can easily observe that the amount of information contained in the world around the entire London underground system must be much more than the information contained in the digital data describing the system.
The latter is much more than the information depicted in the deformed map.
By the time when the workflow reaches a task, the number of decision options is usually limited. For example, counting the number stops may have optional values between 0 and 50.
The amount of information contained in the counting result is much smaller than that in the deformed map.
This evidences the general trend observed in \cite{Chen:2016:TVCG}.

After considering the positive contribution of information, we must counterbalance AC by the the term \emph{Potential Distortion} (PD), which describes, in abstract, the negative consequences that may be caused by information loss. In the past, one typically uses a third-party metric to determine whether a chosen decision option is good or not. This introduces a dilemma that one needs a fourth-party metric to determine if the third-party metric is good or not, and this can go on forever. At least, mathematically, this unbounded reasoning paradigm is undesirable.
This third-party metric was avoided in Eq.\,\ref{eq:CBM-1a} by imagining if a viewer would have to reconstruct the original data that is visualized, how much the reconstructed data would diverge from the original data.
In \cite{Chen:2016:TVCG}, this divergence is measured using the well-known Kullback-Leibler divergence (KL-divergence) \cite{Kullback:1951:AMS}.
Because this divergence measure is unbounded, Chen and Sbert proposed to replace it with bounded measure in a paper that precedes this work and is included in the supplementary materials. We will detail the concerns about the unboundedness in the next section.

As shown in Eq.\,\ref{eq:CBM-1a}, the AC term makes a positive contribution, the DC term makes a negative contribution, reflecting the two sides of the same coin of information loss. Meanwhile, both terms have the same unit \emph{bit}, and are moderated by the term \emph{Cost}.
The term AC characterizes many useful approaches in visualization and visual analytics, such as data filtering and visual abstraction, while the term PD characterizes many undesirable shortcomings such as rendering errors and perceptual errors.   
The term \emph{Cost} encompasses all costs of the visualization process, including computational costs (e.g., visual mapping and rendering), cognitive costs (e.g., cognitive load), and consequential costs (e.g., impact of errors).
The term is defined with as an energy measure, but can be approximated using time, monetary, and other appropriate measures.

The cost-benefit ratio in Eq.\,\ref{eq:CBM-1a} can also be used to measure other processes in a visual analytics workflow.
One can simply imagine replacing the block (a$_1$) in Figure \ref{fig:Concept} with one of the other four blocks on the left, (a$_2$) for faithful visual mapping, (b) for statistics, (c) for algorithms, and (d) for interactive information retrieval.
This exercise allows us to compare the relative merits among the four major components of visual analytics, i.e., statistics, algorithms, visualization, and interaction \cite{Chen:2011:C}.

For example, statistics may be able to deliver a set of indicators about the London underground map to a user. In comparison with the deformed map, these statistical indicators contain much less information than the map, offering more AC contribution.
Meanwhile, if a user is asked to imagine how the London underground system looks like, having these statistical indicators will not be very helpful.
Hence statistics may cause more PD.

Of course, whether to use statistics or visualization may be task-dependent.
Mathematically, this is largely determined by both the PD and \emph{Cost} associated with the perception and cognition process in Figure \ref{fig:Concept}.
If a user tries to answer a statistical question using the visualization, it is likely to cost more using statistics, provided that the statistical answer has already been computed or statistical calculation can be performed easily and quickly.

Whether to use statistics or visualization may also be user-dependent.
Consider a user \textbf{A} has a fair amount of prior knowledge about the London underground system, but another user \textbf{B} has little.
If both are shown some statistics about the system (e.g., the total number of stations of each line), \textbf{A} can redraw the deformed map more accurately than \textbf{B} and more accurately than without the statistics, even though the statistical information is not meant to support the users' this task.
Hence to \textbf{A}, having a deformed map to help appreciate the statistics may not be necessary, while to \textbf{B}, viewing both statistics and the deformed map may help reduced the PD but may also incur more cost in terms of effort.
Hence visualization is more useful to \textbf{B}.

This example echos the scenario presented in Figure \ref{fig:InfoLoss}, where we asked two questions:  Can information theory explain this phenomenon? Can we quantitatively measure some factors in this visualization process?
If prior knowledge can explain the trade-off among AC, PD, and \emph{Cost} in comparing statistics and deformed map.
We can also extrapolate this reasoning to analyze the trade-off in comparing viewing the deformed map (more AC) and viewing the faithful map (less AC) as in Figure \ref{fig:InfoLoss}. 
Perhaps we can now be more confident to say that information theory can explain such a phenomenon.
In the remainder of the paper, we will demonstrate the potential answer to the second question, i.e., we can quantitatively measure some relevant factors in such a visualization process.

To some readers, it may still be counter-intuitive to consider that information loss has a positive side. It is essential for asserting why visualization is useful as well as asserting the usefulness of statistics, algorithms, and interaction since they all usually cause information loss \cite{Chen:2019:CGF}.
Further discourse on this topic can be found in Appendix \ref{app:OriginalTheory}.

% --------------------
\begin{table*}[t]
  \centering
  \caption{The answers by ten surveyees to the questions in the volume visualization survey. The surveyees are ordered from left to right according to their own ranking about their knowledge of volume visualization. Correct answers are indicated by letters in brackets. The upper case letters (always in brackets) are the most appropriate answers, while the lower case letters with brackets are acceptable answers as they are correct in some circumstances. The lower case letters without brackets are incorrect answers.}
  \begin{tabular}{@{}l|cccccccccc@{}}
  & \multicolumn{8}{c}{\textbf{Surveyee's ID}}\\
 \textbf{Questions (with correct answers in brackets)}
     & S1 & S2 & S3 & S4 & S5 & S6 & S7 & S8 & P9 & P10\\
  \hline
  1. Use of different transfer functions (D), dataset: Carp
     & (D) & (D) & (D) & (D) & (D) & c & b & (D) & a & c\\
  2. Use of translucency in volume rendering (C), dataset: Engine Block
     & (C) & (C) & (C) & (C) & (C) & (C) & (C) & (C) & d & (C)\\
  3. Omission of voxels of soft tissue and muscle (D), dataset: CT head
     & (D) & (D) & (D) & (D) & b & b & a & (D) & a & (D)\\
  4. sharp objects in volume-rendered CT data (C), dataset: CT head
     & (C) & (C) & a & (C) & a & b & d & b & b & b\\ 
  5. Loss of 3D information with MIP (B, a), dataset: Aneurism
     & (a) & (B) & (a) & (a) & (a) & (a) & D & (a) & (a) & (a)\\
  6. Use of volume deformation (A), dataset: CT head
     & (A) & (A) & b & (A) & (A) & b & b & (A) & b & b\\
  7. Toe nails in non-photo-realistic volume rendering (B, c): dataset: Foot
     & (c) & (c) & (c) & (B) & (c) & (B) & (B) & (B) & (B) & (c)\\
  8. Noise in non-photo-realistic volume rendering (B): dataset: Foot
     & (B) & (B) & (B) & (B) & (B) & (B) & a & (B) & c & (B)\\
  \hline
  9. Knowledge about 3D medical imaging technology [1 lowest. 5 highest]
     & 4 & 3 & 4 & 5 & 3 & 3 & 3 & 3 & 2 & 1\\
  10. Knowledge about volume rendering techniques [1 lowest. 5 highest]
     & 5 & 5 & 4-5 & 4 & 4 & 3 & 3 & 3 & 2 & 1\\
  \hline
  \end{tabular}
  \label{tab:VolVis}
  \vspace{-2mm}
\end{table*}

% =================================================
\section{\textbf{Formulae of the Basic and Relevant Information-Theoretic Measures}}
\label{app:InfoTheory}
This section is included for self-containment. Some readers who have the essential knowledge of probability theory but are unfamiliar with information theory may find these formulas useful.

Let $\mathbb{Z} = \{ z_1, z_2, \ldots, z_n \}$ be an alphabet and $z_i$ be one of its letters.
$\mathbb{Z}$ is associated with a probability distribution or probability mass function (PMF) $P(\mathbb{Z}) = \{ p_1, p_2, \ldots, p_n \}$ such that
$p_i = p(z_i) \ge 0$ and $\sum_{1}^n p_i = 1$. The \textbf{Shannon Entropy} of $\mathbb{Z}$ is:

\[
  \mathcal{H}(\mathbb{Z}) = \mathcal{H}(P)= - \sum_{i=1}^n p_i \log_2 p_i \quad \text{(unit: bit)}
\]

Here we use base 2 logarithm as the unit of bit is more intuitive in context of computer science and data science.

An alphabet $\mathbb{Z}$ may have different PMFs in different conditions.
Let $P$ and $Q$ be such PMFs. The \textbf{KL-Divergence} $\mathcal{D}_{KL}(P||Q)$ describes the difference between the two PMFs in bits:
\[
  \mathcal{D}_{KL}(P||Q) = \sum_{i=1}^n p_i \log_2 \frac{p_i}{q_i} \quad \text{(unit: bit)}
\]
$\mathcal{D}_{KL}(P||Q)$ is referred as the divergence of $P$ from $Q$.
This is not a metric since $\mathcal{D}_{KL}(P||Q) \equiv \mathcal{D}_{KL}(Q||P)$ cannot be assured.

Related to the above two measures, \textbf{Cross Entropy} is defined as:
\[
  \mathcal{H}(P, Q) = \mathcal{H}(P) + \mathcal{D}_{KL}(P||Q) = - \sum_{i=1}^n p_i \log_2 q_i \quad \text{(unit: bit)}
\]
Sometimes, one may consider $\mathbb{Z}$ as two alphabets $\mathbb{Z}_a$ and $\mathbb{Z}_b$ with the same ordered set of letters but two different PMFs.
In such case, one may denote the KL-Divergence as $\mathcal{D}_{KL}(\mathbb{Z}_a||\mathbb{Z}_b)$, and the cross entropy as $\mathcal{H}(\mathbb{Z}_a, \mathbb{Z}_b)$.

% =================================================
\section{\textbf{Survey Results of Useful Knowledge in Volume Visualization}}
\label{app:VolVis}
This survey consists of eight questions presented as slides.
The questionnaire is given as part of the supplementary materials. 
The ten surveyees are primarily colleagues from the UK, Spain, and the USA.
They include doctors and experts of medical imaging and visualization, as well as several persons who are not familiar with the technologies of medical imaging and data visualization.
Table \ref{tab:VolVis} summarizes the answers from these ten surveyees.

There is also a late-returned survey form that was not included in the analysis. As a record, the answers in this survey form are:
1: c,
2: d,
3: (D),
4: a,
5: (a),
6: (A),
7: (c),
8: (B),
9: 5,
10: 4.
The upper case letters (always in brackets) are the most appropriate answers, while the lower case letters with brackets are acceptable answers as they are correct in some circumstances. The lower case letters without brackets are incorrect answers.

%\clearpage

\begin{table}[t]
  \centering
  \caption{The answers by four surveyees at University of Oxford to the questions in the London underground survey.}
  %\scalebox{1}{
  \begin{tabular}{@{}l@{}r@{\hspace{5mm}}c@{\hspace{3mm}}c@{\hspace{3mm}}%
  c@{\hspace{3mm}}c@{\hspace{5mm}}r@{}}
  & & \multicolumn{4}{c}{\textbf{Surveyee's ID}}\\
  \multicolumn{2}{@{}l}{\textbf{Questions}} 
     & P13 & P14 & P15 & P16 & mean\\
  \hline
  Q1: & answer (min.) & 15 & 20 & 15 & 15 & 16.25 \\
      & time (sec.) & 11.81 & 18.52 & 08.18 & 07.63 & 11.52 \\
  \hline
  Q2: & answer (min.) & 5 & 5 & 15 & 15 & 10.00\\
      & time (sec.) & 11.10 & 02.46 & 13.77 & 10.94 & 09.57 \\
  \hline
  Q3: & answer (min.) & 35 & 60 & 30 & 25 & 37.50 \\
      & time (sec.) & 21.91 & 16.11 & 10.08 & 22.53 & 17.66 \\
  \hline
  Q4: & answer (min.) & 20 & 30 & 60 & 25 & 33.75\\
      & time (sec.) & 13.28 & 16.21 & 08.71 & 18.87 & 14.27 \\
  \hline
  Q5: & time 1 (sec.) & 17.72 & 07.35 & 17.22 & 09.25 & 12.89 \\
      & time 2 (sec.) & 21.06 & 17.00 & 19.04 & 12.37 & 17.37\\
      & answer (10) & 10 & 8 & 10 & 10 & \\
      & time (sec.) & 04.82 & 02.45 & 02.96 & 15.57 & 06.45\\
  \hline
  Q6: & time 1 (sec.) & 35.04 & 38.12 & 11.29 & 07.55 & 23.00 \\
      & time 2 (sec.) & 45.60 & 41.32 & 20.23 & 40.12 & 36.82 \\
      & answer (9) & 9 & 10 & 9 & 8 & \\
      & time (sec.) & 03.82 & 13.57 & 08.15 & 34.32 & 14.97 \\
  \hline
  Q7: & time 1 (sec.) & 01.05 & 02.39 & 09.55 & 11.19 & 06.05 \\
      & time 2 (sec.) & 02.15 & 05.45 & 09.58 & 13.47 & 07.66 \\
      & answer (7) & 10 & 6 & 7 & 7 & \\
      & time (sec.) & 01.06 & 01.60 & 02.51 & 14.06 & 04.81 \\ 
  \hline
  Q8: & time 1 (sec.) & 08.74 & 26.14 & 20.37 & 15.01 & 17.57 \\
      & time 2 (sec.) & 16.50 & 30.55 & 27.01 & 17.91 & 22.99 \\
      & answer (6) & 6 & 6 & 6 & 6 \\
      & time (sec.) & 09.30 & 03.00 & 02.11 & 04.94 & 04.48 \\
  \hline
  Q9: & answer (P) & P & P & P & P & \\
      & time (sec.) & 05.96 & 09.38 & 04.56 & 05.16 & 06.27\\
  \hline
  Q10: & answer (LB) & LB & LB & LB & LB & \\
      & time (sec.) & 12.74 & 07.77 & 01.30 & 09.94 & 07.94 \\
  \hline
  Q11: & answer (WP) & WP & WP & WP & WP & \\
      & time (sec.) & 09.84 & 04.43 & 03.39 & 07.18 & 06.21 \\
  \hline
  Q12: & answer (FP) & FP & FP & FP & FP & \\
      & time (sec.) & 06.22 & 10.46 & 06.78 & 05.10 & 07.14 \\
  \hline
  \multicolumn{2}{@{}l}{live in metro city} & never & days & days & days \\ 
  \multicolumn{2}{@{}l}{live in London} & never & days & days & days \\ 
  \hline
  \end{tabular}
  %}
  \label{tab:MapStudyOU}
\end{table}

% =================================================
\section{\textbf{Survey Results of Useful Knowledge in Viewing London Underground Maps}}
\label{app:London}

\begin{figure}[t]
    \centering
    \includegraphics[width=\linewidth]{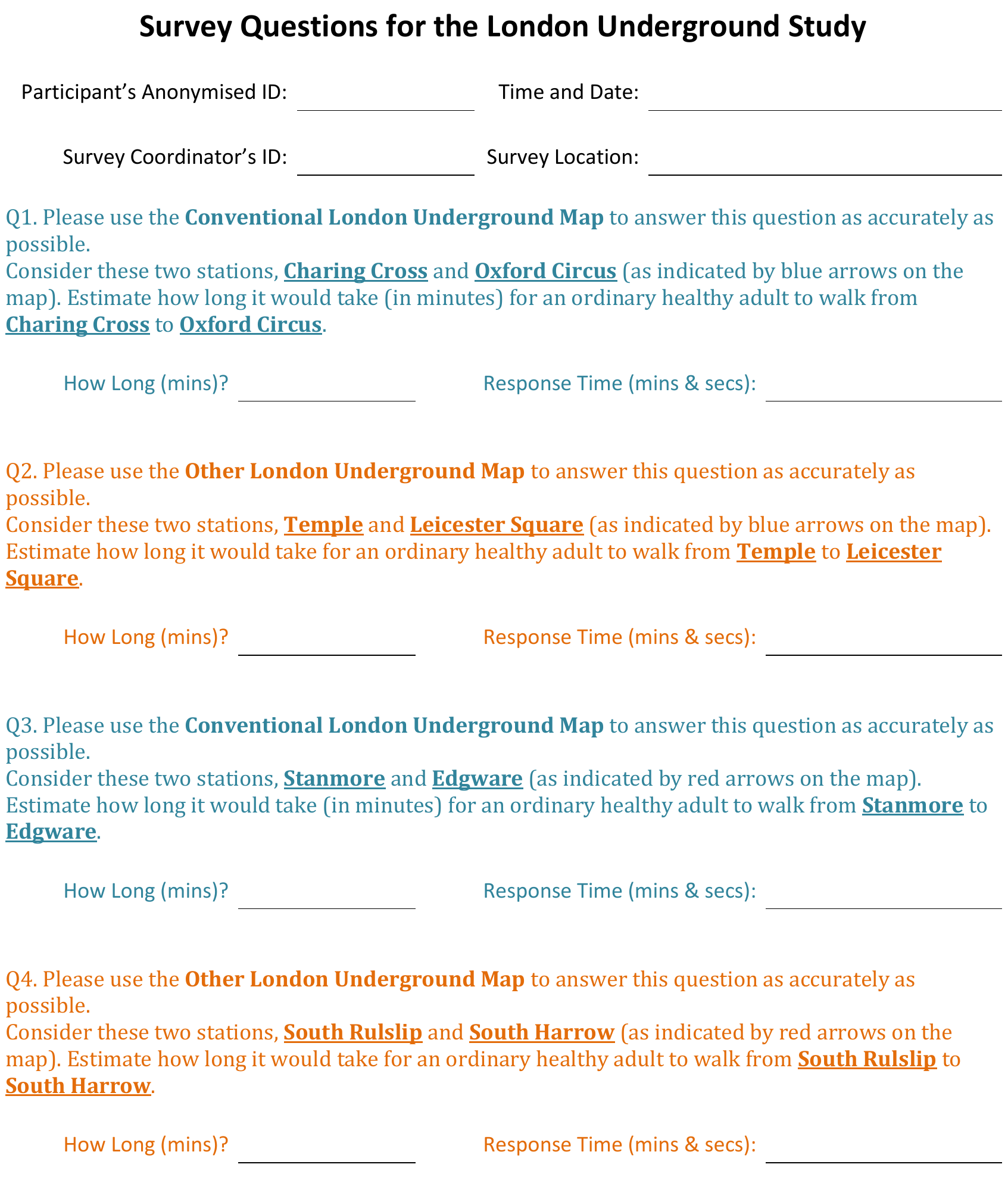}
    \caption{London underground survey: question sheet 1 (out of 3).}
    \label{fig:LondonSheet1}
    \vspace{-4mm}
\end{figure}

\begin{figure}[t]
    \centering
    \includegraphics[width=\linewidth]{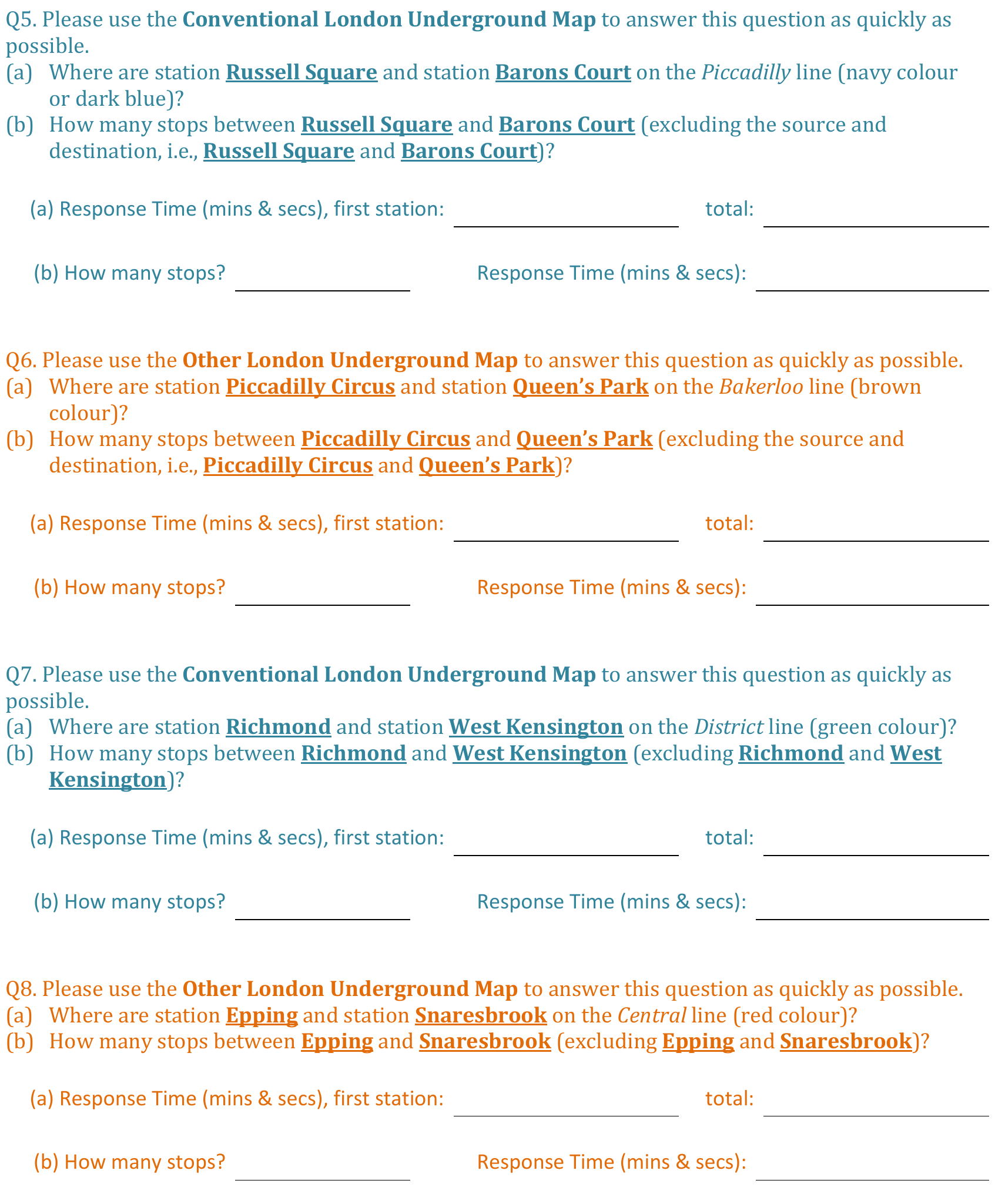}
    \caption{London underground survey: question sheet 2 (out of 3).}
    \label{fig:LondonSheet2}
\end{figure}

\begin{figure}[!ht]
    \centering
    \includegraphics[width=\linewidth]{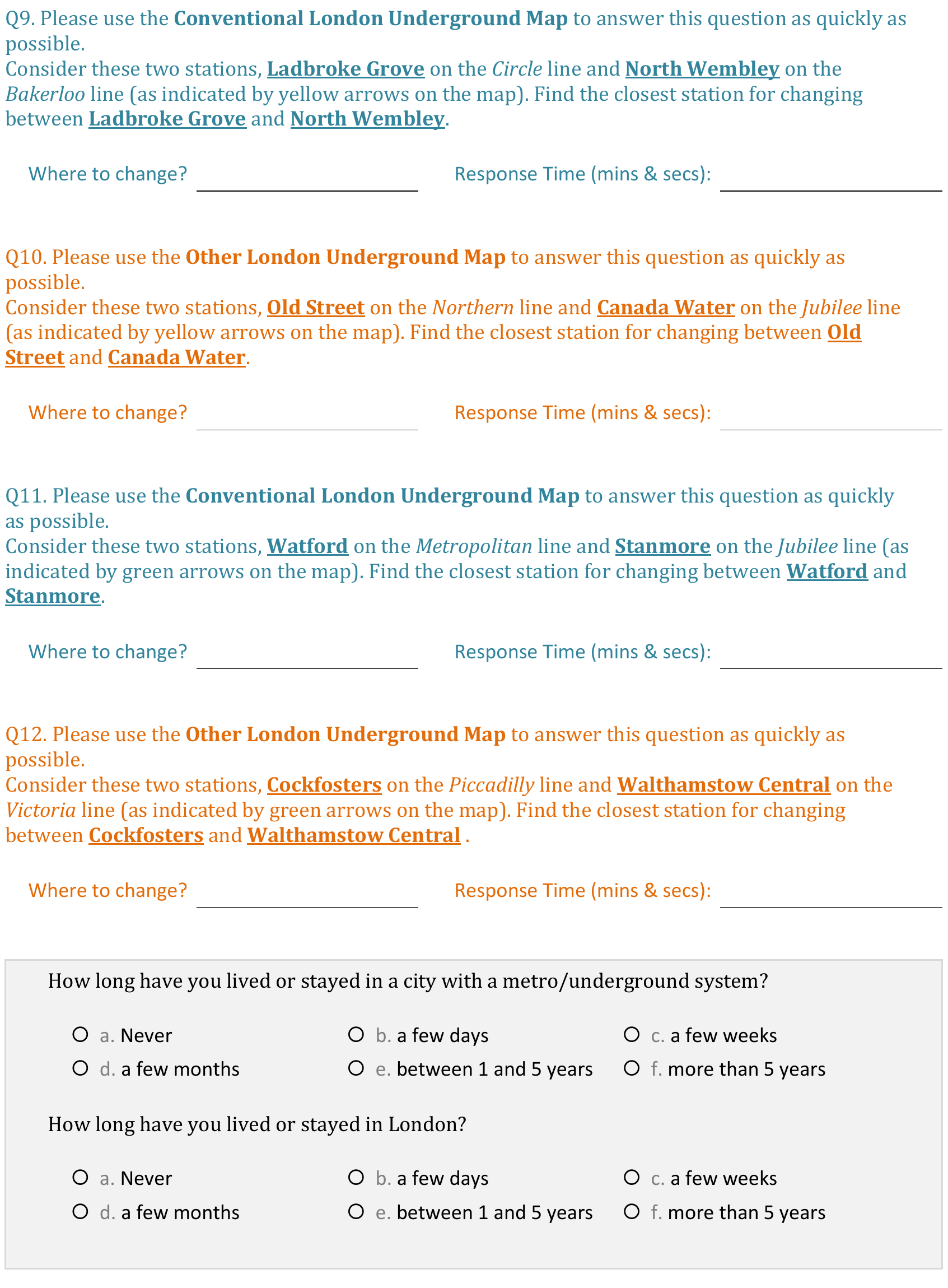}
    \vspace{-4mm}
    \caption{London underground survey: question sheet 3 (out of 3).}
    \label{fig:LondonSheet3}
    \vspace{-4mm}
\end{figure}

Figures \ref{fig:LondonSheet1}, \ref{fig:LondonSheet2}, and \ref{fig:LondonSheet3} show the questionnaire used in the survey about two types of London Underground maps.
Table \ref{tab:MapStudyKCL} summarizes the data from the answers by the 12 surveyees at King's College London, while Table \ref{tab:MapStudyOU} summarizes the data from the answers by the four surveyees at University Oxford.

In Section \ref{sec:London}, we have discussed Questions 1$\sim$4 in some detail.
In the survey, Questions 5$\sim$8 constitute the second set.   
Each question asks surveyees to first identify two stations along a given underground line, and then determine how many stops between the two stations.
All surveyees identified the stations correctly for all four questions, and most have also counted the stops correctly.
In general, for each of these cases, one can establish an alphabet of all possible answers in a way similar to the example of walking distances.
However, we have not observed any interesting correlation between the correctness and the surveyees' knowledge about metro systems or London.

With the third set of four questions, each questions asks surveyees to identify the closest station for changing between two given stations on different lines.
All surveyees identified the changing stations correctly for all questions.

\begin{figure}[ht]
  \centering
  \includegraphics[width=80mm]{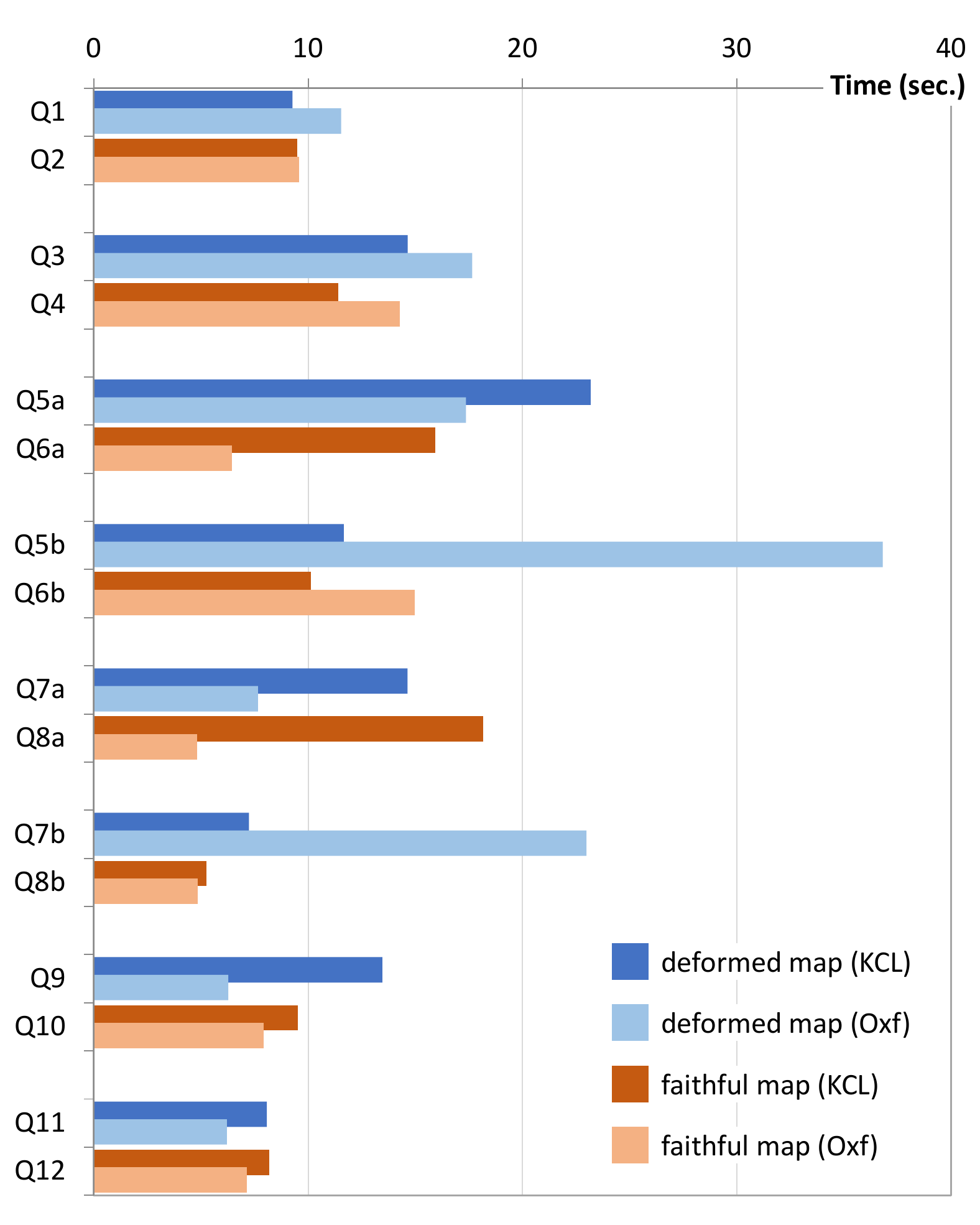}
  \vspace{-2mm}
  \caption{The average time used by surveyees for answering each of the 12 questions. The data does not indicate any significant advantage of using the geographically-deformed map.}
  \label{fig:MapStudy-RT}
  \vspace{-4mm}
\end{figure}

The design of Questions 5$\sim$12 was also intended to collect data that might differentiate the deformed map from the faithful map in terms of the time required for answering questions.
As shown in Figure \ref{fig:MapStudy-RT}, the questions were paired, such that the two questions feature the same level of difficulties.
Although the comparison seems to suggest that the faithful map might have some advantage in the setting of this survey, we cannot be certain about this observation as the sample size is not large enough.
In general, we cannot draw any meaningful conclusion about the cost in terms of time.
We hope to collect more real world data about the timing cost of visualization processes for making further advances in applying information theory to visualization.

Meanwhile, we consider that the space cost is valid consideration.
While both maps have a similar size (i.e., deformed map: 850mm$\times$580mm, faithful map: 840mm$\times$595mm), their font sizes for station labels are very different.
For long station names, ``High Street Kensington'' and ``Totteridge \& Whetstone'', the labels on the deformed map are of 35mm and 37mm in length, while those on the faithful map are of 17mm and 18mm long.
Taking the height into account, the space used for station labels in the deformed map is about four times of that in the faithful map.
In other worlds, if the faithful map were to display its labels with the same font size, the cost of the space would be four times of that of the deformed map.

\clearpage

\begin{table*}[t]
  \centering
  \caption{The answers by twelve surveyees at King's College London to the questions in the London underground survey.}
  \scalebox{1}{
  \begin{tabular}{@{}l@{}r@{\hspace{5mm}}c@{\hspace{3mm}}c@{\hspace{3mm}}c@{\hspace{3mm}}%
  c@{\hspace{3mm}}c@{\hspace{3mm}}c@{\hspace{3mm}}c@{\hspace{3mm}}c@{\hspace{3mm}}%
  c@{\hspace{3mm}}c@{\hspace{3mm}}c@{\hspace{3mm}}c@{\hspace{5mm}}r@{}}
  & & \multicolumn{12}{c}{\textbf{Surveyee's ID}}\\
  \multicolumn{2}{@{}l}{\textbf{Questions}} 
     & P1 & P2 & P3 & P4 & P5 & P6 & P7 & P8 & P9 & P10 & P11 & P12 & mean\\
  \hline
  Q1: & answer (min.) & 8 & 30 & 12 & 16 & 20 & 15 & 10 & 30
      & 20 & 20 & 20 & 30 & 19.25\\
      & time (sec.) & 06.22 & 07.66 & 09.78 & 11.66 & 03.72 & 04.85 & 08.85 & 21.12
      & 12.72 & 11.22 & 03.38 & 10.06 & 09.27\\
  \hline
  Q2: & answer (min.) & 15 & 30 & 5 & 22 & 15 & 14 & 20 & 20 
      & 25 & 25 & 25 & 20 & 19.67\\
      & time (sec.) & 10.25 & 09.78 & 06.44 & 09.29 & 12.12 & 06.09 & 17.28 & 06.75
      & 12.31 & 06.85 & 06.03 & 10.56 & 09.48\\
  \hline
  Q3: & answer (min.) & 20 & 45 & 10 & 70 & 20 & 20 & 20 & 35
      & 25 & 30 & 20 & 240 & 46.25\\
      & time (sec.) & 19.43 & 13.37 & 10.06 & 09.25 & 14.06 & 10.84 & 12.46 & 19.03
      & 11.50 & 16.09 & 11.28 & 28.41 & 14.65\\
  \hline
  Q4: & answer (min.) & 60 & 60 & 35 & 100 & 30 & 20 & 45 & 35
      & 45 & 120 & 40 & 120 & 59.17\\
      & time (sec.) & 11.31 & 10.62 & 10.56 & 12.47 & 08.21 & 07.15 & 18.72 & 08.91
      & 08.06 & 12.62 & 03.88 & 24.19 & 11.39\\
  \hline
  Q5: & time 1 (sec.) & 22.15 & 01.75 & 07.25 & 03.78 & 14.25 & 37.68 & 06.63 & 13.75
      & 19.41 & 06.47 & 03.41 & 34.97 & 14.29\\
      & time 2 (sec.) & 24.22 & 08.28 & 17.94 & 05.60 & 17.94 & 57.99 & 21.76 & 20.50
      & 27.16 & 13.24 & 22.66 & 40.88 & 23.18\\
      & answer (10) & 10 & 10 & 10 & 9 & 10 & 10 & 10 & 10
      & 9 & 10 & 10 & 10 \\
      & time (sec.) & 06.13 & 28.81 & 08.35 & 06.22 & 09.06 & 06.35 & 09.93 & 12.69
      & 10.47 & 05.54 & 08.66 & 27.75 & 11.66\\
  \hline
  Q6: & time 1 (sec.) & 02.43 & 08.28 & 01.97 & 08.87 & 05.06 & 02.84 & 06.97 & 10.15
      & 18.10 & 21.53 & 03.00 & 07.40 & 08.05\\
      & time 2 (sec.) & 12.99 & 27.69 & 04.81 & 10.31 & 15.97 & 04.65 & 17.56 & 16.31
      & 20.25 & 24.69 & 15.34 & 20.68 & 15.94\\
      & answer (9) & 9 & 10 & 9 & 9 & 4 & 9 & 9 & 9
      & 8 & 9 & 9 & 9\\
      & time (sec.) & 07.50 & 06.53 & 04.44 & 16.53 & 19.41 & 05.06 & 13.47 & 07.03
      & 12.44 & 04.78 & 07.91 & 16.34 & 10.12\\
  \hline
  Q7: & time 1 (sec.) & 17.37 & 08.56 & 01.34 & 03.16 & 08.12 & 01.25 & 21.75 & 15.56
      & 02.81 & 07.84 & 02.22 & 46.72 & 11.39\\
      & time 2 (sec.) & 17.38 & 13.15 & 02.34 & 03.70 & 08.81 & 02.25 & 22.75 & 26.00
      & 17.97 & 10.37 & 03.18 & 47.75 & 14.64\\
      & answer (7) & 7 & 7 & 7 & 7 & 6 & 7 & 7 & 7
      & 6 & 7 & 7 & 7\\
      & time (sec.) & 07.53 & 06.34 & 03.47 & 03.87 & 02.75 & 04.09 & 02.16 & 04.94
      & 26.88 & 05.31 & 06.63 & 12.84 & 07.23\\ 
  \hline
  Q8: & time 1 (sec.) & 12.00 & 08.50 & 06.09 & 02.88 & 08.62 & 14.78 & 19.12 & 08.53
      & 12.50 & 10.22 & 12.50 & 20.00 & 11.31\\
      & time 2 (sec.) & 13.44 & 10.78 & 23.37 & 09.29 & 13.03 & 36.34 & 23.55 & 09.50
      & 13.53 & 10.23 & 32.44 & 22.60 & 18.18\\
      & answer (6) & 6 & 6 & 6 & 6 & 6 & 6 & 6 & 6
      & 6 & 6 & 6 & 6\\
      & time (sec.) & 02.62 & 05.94 & 02.15 & 04.09 & 04.94 & 07.06 & 07.50 & 04.90
      & 04.37 & 04.53 & 05.47 & 09.43 & 05.25\\
  \hline
  Q9: & answer (P) & P & P & P & P & P & P & P & P
      & P & P & P & P & \\
      & time (sec.) & 35.78 & 02.87 & 07.40 & 13.03 & 06.97 & 52.15 & 13.56 & 02.16
      & 08.13 & 09.06 & 01.93 & 08.44 & 13.46\\
  \hline
  Q10: & answer (LB) & LB & LB & LB & LB & LB & LB & LB & LB
      & LB & LB & LB & LB & \\
      & time (sec.) & 05.50 & 03.13 & 12.04 & 14.97 & 07.00 & 26.38 & 11.31 & 03.38
      & 06.75 & 07.47 & 06.50 & 09.82 & 09.52\\
  \hline
  Q11: & answer (WP) & WP & WP & WP & WP & WP & WP & WP &WP
      & WP & WP & WP & WP & \\
      & time (sec.) & 06.07 & 05.35 & 07.72 & 05.00 & 04.32 & 23.72 & 05.25 & 03.07
      & 10.66 & 05.37 & 02.94 & 17.37 & 08.07\\
  \hline
  Q12: & answer (FP) & FP & FP & FP & FP & FP & FP & FP & FP
      & FP & FP & FP & FP & \\
      & time (sec.) & 05.16 & 02.56 & 11.78 & 08.62 & 03.60 & 19.72 & 11.28 & 03.94
      & 20.72 & 01.56 & 02.50 & 06.84 & 08.19\\
  \hline
  \multicolumn{2}{@{}l}{live in metro city} & $>$5yr & $>$5yr & mths & 1-5yr
    & $>$5yr & 1-5yr & weeks & $>$5yr & 1-5yr & $>$5yr & mths & mths\\ 
  \multicolumn{2}{@{}l}{live in London} & $>$5yr & $>$5yr & mths & 1-5yr
    & 1-5yr & mths & mths & mths & mths & mths & mths & mths\\ 
  \hline
  \end{tabular}
  }
  \label{tab:MapStudyKCL}
\end{table*}
\clearpage

% =================================================
\section{\textbf{Authors' Revision Report following the SciVis 2020 Reviews}}
\label{app:Feedback}

2 December 2020

\noindent Dear EuroVis 2021 co-chairs, IPC members, and Reviewers,

\noindent\textbf{Bounded Measure for Estimating the Benefit of Visualization: Case Studies and Empirical Evaluation\\
Min Chen, Alfie Abdul-Rahman, Deborah Silver, and Mateu Sbert}

The original version of this paper, ``1051: A Bounded Measure for Estimating the Benefit of Visualization, by Chen, Sbert, Abdul-Rahman, and Silver'', was submitted to EuroVis 2020, the paper received scores (5, 4.5, 2.5, 2). The co-chairs recommended for the paper to be resubmitted to CGF after a major revision (i.e., fast track). Unfortunately, the first author was an editor-in-chief of CGF until January 2020 and still has administrative access to the CGF review system at that time. Hence it will not be possible for CGF to arrange a confidential review process for this paper.

The paper was revised according to the reviewers' comments and was subsequently submitted to SciVis 2020. The SciVis submission, the revision report, and the EuroVis 2020 reviews can be found as an arXiv report https://arxiv.org/pdf/2002.05282.pdf.

The SciVis 2020 submission improved upon the EuroVis 2020 with additional clarifications in the main text and detailed explanations in appendices including a mathematical proof. The reviewers were not impressed and scored the submission as [3.5 2, 2, 1.5], which is much lower than the scores for the EuroVis 2020 submission. We appreciate that all four SciVis 2020 reviewers stated transparently that they had (and likely still have) doubts about the theoretic measure proposed by Chen and Golan, TVCG, 2016, which this paper tries to improve. We also appreciate that the reviewers stated that they tried not to let these doubts to prejudice this work.

We notice that almost all comments that have influenced the rejection decision are in the form of requiring further explanations and clarifications. As the summary review states ``The paper demands a lot from the reader; without detailed knowledge of information theory concepts, as well as details of the cost-benefit ratio, ...'', clearly simply addressing this issue through adding appendices has not met the reviewers' concerns. When we converted the SciVis 2020 submission into the CGF format, it resulted in 11.5 pages, 1.5 pages above the limit of EuroVis 2021 guideline. Considering the needs (i) to add a fair amount of additional explanation and clarification, and (ii) to reduce the paper length to 10 pages, we decided to split our paper into two papers. We are aware that the EuroVis and CGF rarely see two-part publications. We hence rewrite the two submissions substantially, making them self-contained and relatively independent, while avoiding unnecessary duplications. This is the second paper on ``Case Studies and Empirical Evaluation''. The first paper, entitled

\noindent\emph{Bounded Measure for Estimating the Benefit of Visualization: Theoretical Discourse and Conceptual Evaluation\\
Min Chen and Mateu Sbert}

\noindent is included in the supplementary materials. Similarly, this submission is also included the supplementary materials of the first paper. Both submissions will be available as arXiv reports, together with the corresponding revision reports.

In this revision report, we focus on the reviewers' concerns and queries related to the case studies and empirical evaluation in the original EuroVis and SciVis submissions. These comments are highlighted in \textcolor{violet}{purple}. The comments highlighted in \textcolor{orange}{orange} are to be addressed in the first paper.

\noindent{Yours sincerely,\\
Min Chen on behalf of all authors}\\
\rule{4cm}{0.4pt}

\textcolor{violet}{1.	The paper demands a lot from the reader; without detailed knowledge of information theory concepts, as well as details of the cost-benefit ratio, the paper is difficult to understand. It would be better if the basic terms and concepts were briefly presented again in the paper, with reference to the appendix) (R1). It would be nice if some of the two case studies were interwoven into the mathematical presentation to better anchor the concepts (R2). The notation needs to be improved (R1),}

	\textbf{Action.} Because we have now split the original paper into two papers, we have freed extra space. We have added a new section on mathematical notation. We use simple metro maps as examples to illustrate some basic notion of information theory.

\textcolor{violet}{2.	When trying to apply the presented theory to real applications, huge gaps open up (R1, R3, R4); it starts with the fact that it is the goal of almost all visualizations to reduce the amount of information and to transmit only the information that is essential for the respective application situation; and it ends with the fact that the case studies in the paper, on closer inspection, hardly support the goals of the work (R1, R3, R4).}

	\textbf{Clarification and Action.} Although the definition of the ``goal'' to reduce information is not the authors' definition, we understand that some reviewers may introduce this definition to highlight a controversy. The actual ``goal'' was never defined as ``simply to reduce information'' in the Chen and Golan's paper as well as all subsequent papers on the topic. The term that the authors used often is ``trade-off'', i.e., among the needs to
	
	\begin{itemize}
	\item increase alphabet compression (i.e., reduce entropy),
    \item reduce potential distortion, and
    \item reduce cost.
	\end{itemize}
	
	The reviewers comment about ``transmit only the information that is essential for the respective application situation'' is correct. It is commonly mentioned as task-dependent and user-dependent. However, this statement first makes some existing theories conditional, e.g., the three principles proposed by Kindlmann and Scheidegger (2004) now have to depend on ``respective application situations'', such as tasks. In other words, it transforms the question ``whether a visualization must be precise'' to another question ``what precision is needed by an application''. The latter question undermines the former. Theoretically, this commonly-used statement has hinted that some existing postulations or wisdoms in visualization may be over-generalized. In fact, the cost-benefit metric was developed partly to address a similar uncertainty in economics: ``what resolution of time series is most appropriate to an economist?'' asked Amos Golan (co-author of the cost-benefit paper).
	To help readers appreciate that the cost-benefit ratio actually takes the measurement of tasks and users into the consideration for workflow optimization, we have added an extra Appendix to explain this.
	
\textcolor{violet}{3.	A summarizing discussion of the assumptions, limitations and scope of the theory presented is missing (R1, R4).}

	\textbf{Action.} We have added new statements in the Conclusions of this paper.

% =================================================
\section{\textbf{SciVis 2020 Reviews}}
\label{app:SciVis2020}
\setlength{\parindent}{0mm}
\setlength{\parskip}{3pt}
\small
\begin{narrowfont}

We regret to inform you that we are unable to accept your IEEE VIS 2020 SciVis Papers submission:
 
  1055 - A Bounded Measure for Estimating the Benefit of Visualization
 
The reviews are included below. IEEE VIS 2020 SciVis Papers had 125 submissions and we conditionally accepted 32, for a provisional acceptance rate of about 25\%.
 
......
 
----------------------------------------------------------------

Reviewer 1 review

  Paper type

    Theory \& Model

  Expertise

    Expert

  Overall Rating

    <b>2 - Reject</b>\\
    The paper is not ready for publication in SciVis / TVCG.\\
    The work may have some value but the paper requires major revisions or
    additional work that are beyond the scope of the conference review cycle to meet
    the quality standard. Without this I am not going to be able to return a score of
    '4 - Accept'.

  Supplemental Materials

    Acceptable with minor revisions (specify revisions in The Review section)

  Justification

    The paper presents a new measure to assess the benefits of using visualization.
    The measure revises Chen and Golan's cost-benefit ratio by proposing to measure
    distortion differently. The previous measure was unbounded and this issue is
    addressed in the revised measures. The work is evaluated using analysis of two
    case studies where subjects' distortion was assessed through questionnaires and
    compared with the distortion as indicated by the proposed measures.

    Overall, the paper presents a novel idea, but it is not ready for publication. It
    lacks detail on fundamental concepts that were presented in earlier work, but are
    not widely known in the community. There are notable discrepancies to prior work,
    that are never addressed. The premise of the work is dubious, since the unbounded
    measure of distortion indicates that, according to the measure, visualizations can
    be arbitrarily bad, but in practice, where the aim is to maximize benefit, this is
    not a problem, since benefit can be bounded from above in the measure. The rating-
    based (pre-)selection of the proposed alternatives needs more details on the
    rating process to be objective. The suggested alternatives are rather ad-hoc and
    only compared to each other, but never to the previous distortion measure in
    whether they capture distortion and thus benefit more accurately. The appropriate
    validation of a measure is to show that it does measure what it was intended to
    measure. The paper is currently still lacking this and it is questionable, whether
    the authors can provide this validation this within the reviewing cycle.

  The Review

    I will elaborate on the issues mentioned above.

    1. The paper lacks detail on fundamental concepts. This includes both information-
    theoretic concepts as well as details concerning the cost-benefit ratio. Although
    the authors provided a brief introduction in the appendix, this should really be
    in the paper in a brief form. Elaborations can remain in the appendix. Information
    that is crucial to understand Equation (2) is lacking: there is no discussion of
    what Z'$\_$i denotes, not how Z$\_$i, Z$\_$i+1, and Z'$\_$i are related. Note how much the
    paper by Chen and Jänicke 10 years ago elaborated on the basic concepts in
    contrast to the submitted work. The authors may have spent 10 years thinking
    deeply about information theory for visualization and are thus deeply familiar
    with it, but they should respect the needs of the readers, who mostly studied
    something else. Especially, if the authors want a large part of the field of
    visualization to understand and apply the methods they propose.

    2. There are also notable differences to prior work that are glossed over. Chen
    and Golan describe visualization as a sequence of transformations from data to a
    decision and say that the benefit for the whole sequences is H(Z$\_$0) - H(Z$\_$n+1) -
    sum$\_$i=0..n D$\_$KL(Z'$\_$i || Z$\_$i); i.e. the distortion accumulates. In contrast, the
    formula in the current submission comprises just one transition from the "ground
    truth" (which presumably aren't the data) to the decision variable about the
    ground truth. Why do the intermediate distortions no longer show up? And how does
    ground-truth relate to data?

    3. The premise of the work is dubious. The introduction notes that the measure of
    distortion used in the cost-benefit ratio is unbounded. It shows both that it is
    unbounded, as well as that even for simple cases, the distortion can be
    arbitrarily large, even though the information is low, and that this is counter-
    intuitive. However, I will hold, that the unboundedness is not a problem in
    practice. Since H(Z$\_$i) is bounded from above by log |Z$\_$i|, H(Z$\_$i+1) and D$\_$KL(Z'$\_$i
    || Z$\_$i) are bounded from below by 0, the benefit, defined as H(Z$\_$i) - H(Z$\_$i+1) -
    D$\_$KL(Z'$\_$i || Z$\_$i), is bounded from above by log |Z$\_$i| as well. It may not be
    bounded by below, but since the aim is to maximize the cost-benefit ratio, this is
    of no concern. Only maximizing functions that cannot be bounded by above are
    problematic. Note that log-likelihood scores also cannot be bounded from below,
    yet statistics has no reservations in using them for maximum-likelihood
    estimation. A different motivation would be Kulback-Leibler divergence
    overestimates distortion and therefore non-confusing visualizations are
    incorrectly claimed to be confusing by the cost-benefit ratio. Chen an Golan wrote
    that negative benefit indicates confusing visualizations. How well do different
    measures for distortion align with this assessment? For which measures does the
    sign of the benefit measure indicate confusing visualizations? For which measures
    do benefit values correlate better with visualization quality?

    4. It doesn't help the discussion that the paper claims that Kullback-Leibler
    divergence is unbounded, and then provide a "proof" in Section 4.1 that it is
    bounded for particular choices of Q. Since the measure of distortion is applied in
    the paper to relate the ground truth and the decision variable, both of which have
    probability distributions not under our control, the premises of the proof are not
    met and its result is therefore irrelevant. The section should be removed.

    5. The suggested alternatives are rather ad-hoc. While Kullback-Leibler
    divergence, cross entropy have interpretations in data transmission contexts
    (which are even recounted in the paper: the expected number of (excess) bits when
    using code lengths for symbols that would be optimal for a distribution P, when
    the actual symbol properties are given by distribution Q). There is no
    interpretation of the new measures given (in the sense of how to translate values
    for these measures to meaningful natural-language sentences), yet these previous
    measures are declared unintuitive and the new one's intuitive without any
    indication of why. The assessment of the different measures uses a catalogue of
    criteria and values are assigned to each measures, again, rather ad-hoc. There are
    no clear rules, under which conditions a certain number (ranging from 0-5) is
    assigned, nor a discussion of inter-coder consistency (Would different people
    assign the same numbers?). Overall, I do not see that such qualitative assessments
    are helping. See next point. [Note that Shannon put forth axioms that a measure
    for uncertainty should have and showed that entropy is the only formula meeting
    all the properties specified axiomatically. This however is quite different to the
    approach taken by the authors of the present submission.]

    6. \textcolor{violet}{The proper validation of a measure is to show that it measures what it is
    intended to measure.} Although there are numerous studies in visualization that
    establish one design as working or working better with respect to some task, the
    cost-benefit ratio was not applied to any of them. Instead the authors performed
    two studies. However, they do not fully model these studies in their framework.
    There is no mention of the sample space underlying the experiment, or the random
    variables. How can we estimate the probability distribution for the decision
    variable from the responses? The authors instead use one probability distribution
    for each answer. Is this meaningful? It removes all the probabilistic effects that
    underlie the foundation of the measures in the first place. Only some outcomes of
    the studies are discussed. Why? The distributions for ground truth are guessed,
    while there is arguably a ground truth present from which the distribution can be
    read off (e.g. the number of subway stations between two given stations). In the
    discussion, the images do not seem to play a role, only the distortion between
    ground truth and response. Furthermore it is confusing that both Q and Q' are
    described as probability distributions for ground truth, that F remains
    mysterious, and that the value for alphabet compression cannot be traced. Do the
    topological and the geographical subway maps have the same alphabet compression?
    Nor is it clear, how Q, Q', and F relate to Z$\_$i, Z$\_$i+1, and Z'$\_$i in Equation (2).
    The distribution function for Q in 5.2 is also ill-justified.

    7. Lastly: notation. The paper uses notation rather freely, e.g. to specify
    probability distributions, which are functions, using set or interval notation,
    which is non-standard, inconsistent, and confusing for people well-versed in
    probability theory. Similar concepts get different symbols, and the same thing is
    referred to by different expressions (e.g. H(Z), H(P(Z)), H(P)). It is also very
    confusing when probability distributions are denoted by their alphabets, in
    particular when different symbols for alphabets refer to probability distributions
    over the same alphabet. The authors should make notation consistent both within
    and between paper and appendix, and use established conventions in probability
    theory and information theory.

----------------------------------------------------------------

Reviewer 2 review

  Paper type

    Theory \& Model

  Expertise

    Knowledgeable

  Overall Rating

    <b>3.5 - Between Possible Accept and Accept</b>

  Supplemental Materials

    Acceptable

  Justification

    The manuscript is an interesting mathematical foray into how loss of information
    can actually benefit a visualization user, and what the bounds of the distortion-
    loss might be. The idea, the analysis, and the proof are valuable contributions to
    the visualization field, in particular since the conclusion is counterintuitive to
    many audiences, and a hangup of some domain experts (those who decline to use
    scientific visualization altogether because of inherent distortions). On the for
    improvement side, the manuscript is not an easy read, and could also benefit from
    certain clarifications. The two real world examples are intriguing, although the
    domain motivation (why these questions) could be further clarified and discussed.

  The Review

    The manuscript contributes a fascinating mathematical foray into how loss of
    information can actually benefit a visualization user. This is a valuable
    contribution to the field, in particular since the conclusion is counterintuitive.
    A major drawback is that the manuscript is not an easy read, in particular for a
    potential computer science or design graduate student, which is a pity, and that
    it could benefit from certain clarifications.

    The manuscript makes several significant contributions, including:
    
    ** a theoretical discussion of information loss, based on a theoretical basis in
    information theory.
    The information theory premise is valid, and builds on peer-reviewed prior
    publications. The same approach gave us, years ago, the first semblance of
    formalism in analyzing Shneiderman's "Overview First" mantra, and putting some
    bounds to it (good for novices, bad for experts). Albeit not perfect, that
    formalism opened up the possibility of a discussion in the field. Information
    Theory is one possible theoretic framework, leading to discussion in the field,
    and so it has merit.

    ** an interesting theoretical discussion of potential distortion (PD), as a term
    that could be calibrated and bounded.
    There seems to be disagreement in the field about whether PD should be absolutely
    minimized, or it could be calibrated as part of a benefit ratio. That is good: it
    means PD is something we need to discuss as a field. This manuscript is the
    beginning of that conversation.

    The PD subway example in the manuscript is brilliant, although somewhat cryptic,
    in that it illustrates well these different views of PD, and how we may be at
    cross-purposes when discussing it: In NYC, the shortest path involves exiting the
    subway system, walking for a few minutes on the street (hence the length judgment,
    and the importance of PD), and re-entering the subway system at a different line.
    In contrast, in the Berlin or Chicago subway, all the lines are well connected,
    and all that matters is the number of stops (PD almost unimportant). However, had
    I not lived in NYC for a few years, I wouldn't even know these things. It's
    possible the authors themselves are unaware of an alternative interpretation of
    subway maps and PD---than the one they used in the paper.

    This may be the case about the manuscript medical vis example, as well---I do not
    have experience educating novice medical students, but the authors do. This fact,
    that the manuscript looks at PD from a different point of view than me, has
    tremendous value to me. It would be helpful to clarify that this manuscript views
    things differently from several of us, and that overall, sometimes PD is just a
    quantity to be minimized, whereas other times PD is a quantity where bounding
    would be useful.

    ** an ad hoc construction of a bound, which the manuscript evaluates via a user
    study.
    Whereas the construction is ad hoc, and there must be other ways of approaching
    the problem, I do not think it detracts significantly from the manuscript. It
    works reasonably well as a proof of concept.

    Overall, this manuscript seems to me to present great opportunities for the sci
    vis field to grow.

    Suggestions for improvement:
    
    * It would be great if some part of the two case studies were weaved into the
    mathematical exposition, to help anchor the concepts. Humans are great at
    generalizing from examples. For example, I really can't parse this statement:
    ``Suppose that Z is actually of PMF P, but is encoded as Eq. 4 based on Q.''
    What does ``based on Q'' mean here?
    The easier to follow the proof and function construction, the better. In general,
    it might be worth moving to the appendix, or compacting the proof and ad hoc
    construction.

    * Instead of reiterating what was done, step by step, in the conclusion, it would
    be good to see some discussion/summary of assumptions, limitations, and where the
    work agrees or disagrees with other published observations.

    * The information theory approach and the PD discussion are interesting, and a
    valid way of approaching this problem. The user study is also a reasonable way to
    evaluate this theory. However, here and in previous manuscripts in this vein, it
    seems to me the supporting evidence is subject to how we interpret the term
    "information".  For example, in the two case studies, the questions asked as not
    typical of the two domains in some settings, but they may be typical in other
    settings. E.g., a clinician would be interested in where an anomaly is located and
    in its shape, and how much uncertainty there is; whereas a nursing student might
    be interested in general literacy of the type described in the manuscript
    evaluation. In the second case study, a Chicago subway user would often be
    interested in the number of stops and line changes, and not so much in walking
    distance, because the lines are well connected; whereas a NYC subway user often
    needs, for the fastest route, to come out of the subway system, walk for 5
    minutes, and enter the subway system at a different point--I see how the type of
    judgments used in the second case study would be useful there. These
    considerations should appear both in the introduction, to help motivate the
    manuscript, and in the discussion, to help anchor it.

    * A reworking of most figure captions would be good (see Minor comments below).

    Minor:
    I am afraid that we are becoming so mathematically illiterate as a field, that the
    manuscript would benefit from reminding the readers what ``unbounded'' means in math
    terms. Also, a gentle brief definition of terms unfamiliar to a graduate student
    (e.g., ``MCDA, a sub-discipline of operations research that explicitly evaluates
    multiple conflicting criteria in decision making'', `` data intelligence workflow,
    *def here*'', ``cross entropy'') might encourage students to keep reading the paper.

    When discussing the relationship between AC and PD, I'd be interested in any
    practical examples supporting the increase/decrease illustrated in Fig.2.

    Citations are grammatically invisible. Not good: ``observational estimation in
    [43]''. Good: ``observational estimation in Smith et al. [43]'' or ``observational
    estimation [43]''.

    Fig 6 and Fig 7
    * Some reorganization of the caption would be helpful (e.g., in Fig 7, I'd start
    with ``Divergence values for six users using five candidate measures, on an example
    scenario with four data values, A, B, C, D. ''
    * ``them'' -> ``the data values'' (?)

    Fig 7 
    * typo: ``A, B, C, are D.'' (and D). 
    * clarify ``pr.'' stands for ``process'', or remove

----------------------------------------------------------------

Reviewer 3 review

  Paper type

    Theory \& Model

  Expertise

    Expert

  Overall Rating

    <b>1.5 - Between Strong Reject and Reject</b>

  Supplemental Materials

    Not acceptable

  Justification

    The basic story of the paper is: information theory is a promising way to
    characterize the process of data vis because it gives ways to quantify information
    transmission. From some of the measures it provides (Shannon entropy, and KL
    divergence), the authors have previously [11] developed a measure of "benefit" (eq
    2).  But now, it is apparently a big problem that the negative term in benefit
    ("potential distortion" or PD) is unbounded, so a bounded measure is needed, and
    the paper aims to make a well-informed pick from among many possible choices.

    However, the authors do not make a case that, insofar as this is about data
    visualization, a bounded measure is actually a problem that merits a whole paper,
    and, the evaluation of the new measure, such as it is, has a strained relationship
    to both information theory and to visualization practice.

    This submission is on a topic of interest to visualization research, and it may be
    enthusiastically received by the adherents of [11] and the related work in the
    same vein. The submission should stand on its own as a piece of scholarship, but
    does not.

  The Review

    The authors are pursuing information theory as a means of putting data vis on
    mathematical footing.  The current submission relies heavily on a previous paper
    [11]. In response to a previous round of reviews (from a rejected Eurovis 2020
    submission), the authors were defensive about critiques that they felt were of
    [11], rather than of the new work, and have written a new summary of [11] work as
    Appendix A.  However, every submission presents a new opportunity for scrutiny,
    from a new set of reviewers, which ideally generates new opportunities for authors
    to strengthen or reconsider a line of work, even if initial steps have passed peer
    review.  Nonetheless, this review will strive to consider the new submission on
    its own terms.

    My high-level considerations are:

    ((1)) the motivation of needing a bounded information theoretic measure of PD, and
    the mathematical strength of the connection to information theory, and

    ((2)) "intuitive" and "practical" (as noted at the end of paragraph 4)
    applicability to vis applications

    First ((1)):

    We accomplish theoretical research when we can reframe new things in terms of pre-
    existing and trusted theory. The authors trust information theory, and advertise,
    in the Abstract and Intro, how they are drawing on information theory.  However,
    the authors seem bothered (top of 2nd column, 1st page) by the fact that the KL
    divergence when communicating a binary variable can be unbounded, because "the
    amount of distortion measured by the KL-divergence often has much more bits than
    the entropy of the information space itself. This is not intuitive to interpret
    .."  Later they say (middle 2nd column page page 3) "it is difficult to imagine
    that the amount of informative [sic?] distortion can be more than the maximum
    amount of information available."

    Actually, it is not difficult to imagine, if you accept information theory (as the
    authors surely want us to do), hence this is a worrisome start to the math
    foundations of this work.

    If you have an extremely biased coin Q (say, p(head)=0.99, p(tail)=0.01), and want
    to use bits to communicate a sequence of flips of Q, the optimal scheme will
    involving something like using "0" to represent some number (greater than 1) of
    heads in a row, i.e., some kind of compression, which can be very efficient for Q
    given how low the entropy is: ~0.081. But then, if you are tasked with encoding a
    sequence of flips of an *unbiased* coin P (p(head) = p(tail) = 0.5), you are going
    to be very disadvantaged with your Q-optimized encoding, so much so that it could
    easily more than double the expected length of the encoding, as compared to a
    P-optimized encoding (i.e. "0"=head "1"=tail). As the authors surely know, the KL
    divergence D$\_$KL(P|Q) tells you the expected number of extra bits needed to
    represent a sample from P, when using a code optimized for Q, rather than one
    optimized for P.  For this specific example, D$\_$KL(P|Q) = 2.32918, which motivates
    how D$\_$KL can naturally exceed 1 even for a binary event.  But again, the authors
    probably know this (in fact their section 4.1 is closely related to this
    situation), so it is unclear what they mean by saying this is "unintuitive".

    Also, if the worry is that KL divergence is unbounded because of the places where
    probabilities can be become very small, then why not try some smoothing?  That is
    a known thing in ML, for example this paper:

    https://papers.nips.cc/paper/8717-when-does-label-smoothing-help.pdf

    talks about "label smoothing" in the context of minimizing cross entropy (which is
    the same as minimizing KL divergence if the original or data distribution P is
    fixed).  So the connection to stats or information theory, as it is commonly used
    in current research, appears weak, which makes the proposed new measure appear
    contrived.

    Another good property of KL divergence is its additivity: P=P1*P2 and Q=Q1*Q2 ==>
    D$\_$KL(P|Q) = D$\_$KL(P1|Q1) + D$\_$KL(P2|Q2), which is relevant for considering how
    models and phenomena can have both coarse-grained and fine-grained components.  KL
    divergence nicely analyzes this in terms of simple addition, which would also seem
    to be valuable for the authors, given how in their driving equation (1) KL
    divergence appears in the numerator of a fraction.

    Instead, the authors invent some new measures (9) and (10), which raises various
    red flags.  Inventing new measures for information divergence seems like an
    interesting research direction, but (red flag \#1) the most informed peer reviewers
    for that will be found in statistics, so it is a little scary for the value of
    these measures to be evaluated by vis researchers, when fundamentally, the
    underlying math actually has nothing to do with visualization (any more than it
    has to do with any other application of statistics).  Red flag \#2 is that the new
    measures involve a free parameter "k", unlike the standard information-theoretic
    measures, but maybe this is okay in the same way that smoothing is okay.  Still,
    it doesn't seem like the new measures will be additive, or this isn't shown (red
    flag \#3).  Red flag \#4 is that section 4.1 ("A Mathematical Proof of Boundedness")
    is not a proof at all - it simply describes one specific scenario where the
    optimal encoding for Q will require more, but a bounded amount more, bit to encode
    a sample of P.  That does not prove anything, nor does it illuminate anything
    about how intuitive or reasonable information divergence measures should behave in
    general. The many red flags undermine the desired rigorous connection to
    information theory.

    If, at the end of the day, there is only a tenuous connection to information
    theory, and yet there is a perceived need to have a bounded information divergence
    measure, why not redirect the creative thought that went into formulating the new
    measures (9) and (10), into some bounded monotonic reparameterization of D$\_$KL,
    like arctan(D$\_$KL)?  The authors have not explained why this much simpler and
    obvious strategy is not acceptable, and yet, based on their motivation from Figure
    2, it should suffice.  Figure 2 nicely documents the desired *qualitative* effects
    of varying AC, PD, and Cost. None of the later quantitative arguments were
    compelling, so it was too bad the authors did not pursue a simpler way of meeting
    their goals that maintains at least some connection to D$\_$KL.

    We finally turn to how the different possible measures are actually compared and
    evaluated.  Criterion 1 of Section 4.3 essentially amounts to looking at lots of
    plots, and making qualitative judgments about what behaves in the desired way.  A
    suitably formulated monotonic reparameterization of D$\_$KL could have done well
    here, and the qualitative nature of this evaluation undermines the author's
    mission to find quantitative measures for visualization.  Criteria 2 through 5 in
    Table 3 involve a 1--5 Likert scale (!) for judging the various alternative
    measures.  Subsequent criteria (described below) are used in a "multi-criteria
    decision analysis (MCDA)" analysis, a fancy way of saying: a ranking based on some
    heuristically selected weighted averages of various desiderata.  The combination
    of Likert scales and MCDA is where we lose any remaining connection to information
    theory.  How we know: a different set of equally plausible candidate measures,
    under a different set of criteria, with a different Likert scales, and different
    weightings, could have produced a different winner.

    \textcolor{orange}{Hence, the chosen measure was not, it turns out, actually determined by
    information theory, and its actual value for *quantitative* analysis was never
    demonstrated.}

    Now for consideration ((2)): "intuitive" and "practical" (as noted at the end of
    paragraph 4) applicability to visualizations

    The exposition starts with the claim that "inaccuracy" is ubiquitous, even
    necessary, in visualization, and that information theory promises to quantify the
    inaccuracy.  Even if this claim was made in [11], it bears scrutiny as the
    foundation of this paper.  As Alfred Korzybski tells us, "the map is not the
    territory" and "the word is not the thing".  OF COURSE there is information
    missing in a visualization; it wouldn't be visualization, and it wouldn't be
    useful for visual analysis and communication, if it merely re-iterated all the
    data.  Visualizations are useful insofar as they make task-specific design
    decisions about WHAT to show (and what not to show), and audience-specific design
    decisions about HOW to show it.  The editing task of choosing what to show is
    where the authors would point out "inaccuracy", but that is not how visualization
    research, or scientific visualization research in particular, understands that
    term, regardless of [11].

    The later evaluation criteria in Sec 4.3 do not convey a realistic consideration
    of visualization applications.  In Criterion 6 the authors embrace an encompassing
    view of an analysis workflow that has to end in some binary decision, in which the
    mapping from data to decision actually has to transmit less than a single bit
    (since p(good)=0.8 so entropy is $\sim$0.72).  If some decision support system were so
    refined and mature that the human user need only made a binary decision, and the
    accuracy of that decision is the entirety of the story, does that seem like a
    compelling setting for visualization research?  There is no room in the analysis
    for using visualization to justify or contextualize the decision, and no down-
    stream stake-holders have any role in the analysis, so why even consider
    visualization, instead of a well-trained machine learning method?  The story for
    Criterion 7 is nearly as contrived, and again not actually admitting any
    visualizations per se, so it is not compelling as a setting for evaluation.

    Vis researchers use case studies with the belief that, even though they assess
    some limited and contrived scenario, they nonetheless reveal something informative
    and representative about broader and more general situations.  The two case
    studies were unconvincing individually, but there is also general problem: where
    are the encodings of the many alphabets Z$\_$i, Z$\_${i+1}, etc, one for each step of
    some workflow?  I thought the authors' theory is based on analyzing each of those
    steps, but the case studies collapse everything from some simplistic reduction of
    the data, to some simplistic reduction of a viewer's response. Because the case
    studies are so structurally disconnected from all the previous mathematical
    exposition, it's harder to see how the case study results support the goals of the
    paper.

    The Volume Visualization (Criterion 8) case study just doesn't work.  The
    questions asked about volume visualization in the questionnaire are completely
    unrecognizable to me, as someone with experience with real-world biomedical
    applications of volume rendering.  The questionnaire involves variations on the
    same false premise- that a good visualization is one that allows you to infer the
    data from the picture. No: in volume visualization the whole point is that there
    are huge families of transformations on the data that are purposely invisible (a
    task will typically require scrutinizing some materials or interfaces and ignoring
    others), and, the questions to answer for the task are not about recovering the
    original data values at any one location in the rendered image, but rather about
    inferring higher-order properties about structures and their geometric inter-
    relationships (e.g. what is the shape of the bone fracture, or, what is the size
    and extent of a lesion relative to other anatomy).  That all the questions here
    involve pre-existing renderings, of *old* test datasets (wholly detached from any
    real-world application), was economical on the part of the authors but deficient
    as scholarship. I am unconvinced that this questionnaire can shed light on how to
    evaluate volume visualizations, or how to  evaluate measures for evaluating volume
    visualizations.

    The London Tube map (Criterion 9) case study is more interesting, but describing
    Beck's map as "deformed" with a "significant loss of information" risks missing
    the point.  Beck recognized that for the commuter's task of navigating the subway,
    absolute geography is irrelevant (because you can't see it), but topology and
    counting stops matter (because that's what you do experience). Beck made a smart
    choice about what to show. Both kinds of maps used in the study are faithful: one
    is faithful to geography (while distorting the simplest depiction of the
    topology), and one is faithful to the topology (while distorting geography). There
    is something interesting to learn from a case study about measuring the mismatch
    between geographic vs topological maps, and geographic vs topological tasks, but
    this paper doesn't convince me that information theory is at all relevant to that.
    Instead, the case study is used here to help choose a bounded information
    distortion measure, but again the reliance on a Likert scale ("spot on", "close",
    "wild guess") is a sign that the ultimate choice of measure is much less connected
    to information theory than the Abstract and Intro imply.

    The more interesting component about the London Tube map case study was that cost
    was considered.  But hang on: why is the choice of a bounded information
    divergence measure -- for the numerator in (1) -- so crucially important, when the
    choice of cost -- for the denominator in (1) -- is basically left completely
    unspecified?  The authors say that cost can be different things "(e.g., in terms
    of energy, time, or money)", but if the goal is meaningful quantitation, how can
    cost be so under-specified?  It could have a huge effect on the relative
    cost/benefit ratios of comparable vis methods, which further undermines the idea
    that the choice of bounded divergence measure alone warrants a paper.

----------------------------------------------------------------

Reviewer 4 review

  Paper type

    Theory \& Model

  Expertise

    Knowledgeable

  Overall Rating

    <b>2 - Reject</b>\\
    The paper is not ready for publication in SciVis / TVCG.\\
    The work may have some value but the paper requires major revisions or
    additional work that are beyond the scope of the conference review cycle to meet
    the quality standard. Without this I am not going to be able to return a score of
    '4 - Accept'.

  Supplemental Materials

    Acceptable

  Justification

    The paper deals with the improvement of a quantitative measure for the cost-
    benefit ratio of visualizations. In previous work (Ref. [11]) a proposal was made
    for this ratio.
    In the counter there are 2 terms, one bounded, the other unbounded. The authors
    argue that both terms should be bounded. The argumentation seems to me not quite
    conclusive, at least it is not clear to me why it is so important.

    Furthermore, the authors are looking for a more suitable mathematical quantity
    that is bounded. They propose some candidates for this; the selection of the most
    suitable quantity seems to me to be rather ad hoc and arbitrary. The main reason
    is that the selection process is not based on realistic examples.

    Correcting these shortcomings seems to me to require a major revision, which is
    not possible within the short revision cycle.

  The Review

    Preliminary remark: 

    I'm convinced that information theory plays an important role in the foundation of
    visualization. However, I believe that information theory based on Shannon's
    mathematical theory of communication (which operates largely on a syntactic level)
    is not sufficient for this. Semantics and task-dependent weighting play too big a
    role in visualization. Although semantics can be represented by suitable
    alphabets, these have to be defined against the background of complex knowledge
    frameworks into which the user sorts the perceived information, utilizing
    perceptual and content-related prior knowledge. Therefore, I think that Shannon's
    theory represents a rigorous constraint to any further theory formation about all
    semantic and pragmatic aspects of information as well as about visualization. The
    question is how strong this constraint is and which dimensions need to be added to
    reflect the importance of semantics in data visualization.

    Furthermore, I think that the benefit of a visualization depends very much on how
    cognitive peculiarities (e.g. the particular strength of processing spatial
    information or the limited capacity of short-term memory) are taken into account
    in the visual design. I do not see (maybe I know too little about) how this is
    taken into account in the information-theoretic framework.

    As I am not an expert in these matters, I do not wish to include my fundamental
    doubts in my assessment. I try to evaluate the paper by asking what it brings to
    the table, *assuming* that the described information theory approach is adequate
    for "estimating the benefit of visualization".

    -----
    Review:

    Why should we be interested in estimating the benefits of visualization
    quantitatively? Of course, because it would help to advance the visualization. So
    the overall goal seems to make sense.

    The paper is based on Ref. [11], in which Chen and Golan proposed an information-
    theoretic measure for the cost-benefit of data visualization workflows. They
    propose to measure the cost-benefit ratio by (alphabet compression - potential
    distortion) / cost. Here, "alphabet compression" is the information loss due to
    visual abstraction and the "potential distortion" is the informative divergence
    between viewing the data through visualization with information loss and reading
    *all* the data.

    The "alphabet compression" is computed as the entropic difference between the
    input and output alphabets. This is a bounded quantity. The "potential distortion"
    was measured in Ref. [11] as the Kullback-Leibler (KL) divergence of an alphabet
    from some reference alphabet. This quantity is unbound. The only goal of the paper
    is to find an appropriate information-theoretical quantity for the "potential
    distortion", which is also *bounded*.

    As a reason for this goal, the authors say that the benefit calculated using the
    KL divergence "is not quite intuitive in an absolute context, and it is difficult
    to imagine that the amount of informative distortion can be more than the maximum
    amount of information available."

    From a practical point of view, I do not really see the problem: as I mentioned
    before, the goal is to improve visualization, which includes trying to keep the
    "potential distortion" small. In other words, you're practically working at the
    other end of the scale and you wouldn't even feel that the quantity is not
    bounded. For example, in the visualization of high-dimensional data, there is an
    entire branch of research that is mainly concerned with finding projections with
    the least possible distortion.

    The fact that the amount of information distortion can be greater than the maximum
    amount of available information could be seen as a theoretical disadvantage. To
    better understand the motivation for the paper, I would like to see this point
    worked out in more detail.

    Some of the statements in the paper puzzled me, especially the repeated statement
    that the following fact would seem counter-intuitive: that high alphabetical
    compression is a feature of a benefit of visualization. This surprises me, as so
    many efforts of visualization are directed in exactly this direction: Information
    reduction, visual abstraction, feature extraction, visual summaries, first
    overview then (user controlled) details... The intuition of visualization experts
    should rather tell them that information reduction is a quality feature - up to
    the point where essential information is no longer visually represented. The last
    point is missing in the model, unless the omission or loss of essential
    information is subsumed under "information distortion".

    I find the search for bounded measures for the Potential Distortion (PD)
    unconvincing. I am aware of the size of the problem: We try to find suitable
    abstractions to create a theoretical framework that covers a huge number of very
    different use cases. This is a mammoth task, which cannot be accomplished in one
    step. Rather, one has to approach it in many attempts with subsequent
    modifications and iterative improvements. Here we want even more, namely a useful
    quantitative formulation that captures essential aspects of information processing
    in the visualization pipeline. One basic problem is the vague imprecise
    terminology, another is the high dimensionality of the problem. I respect any
    attempt to address this huge problem. Nevertheless, the described selection of an
    appropriate mathematical quantity that measures PD I did not find convincing.

    Section 4 starts with the "mathematical proof" (Sect. 4.1). I don't think this is
    a mathematical proof, since a special PMF Q is used here (or is this distribution
    mathematically somehow distinguished?).

    In Sect. 4.2, again a reason is given (the arbitrariness of specifying epsilon)
    why "it is [...] desirable to consider bounded measures that may be used in place
    of D$\_$KL". I don't understand this argumentation.
    In equation (9), it should probably read k $\ge$ 1, so that slope of |...|$\land$k is not
    negative.

    In equation (11), H$\_$max should be explained: over which set is the maximum
    searched for?

    In Sect. 4.3, a search for the "most suitable measure" is compared to the search
    for a suitable unit (e.g., metric vs. imperial). This is a misleading comparison.
    While the mentioned example is about different units of measurement for the same
    operationally defined quantity (units of measurement that can be converted by
    multiplying by a constant factor), this paper is about selecting a suitable
    (operationally defined) quantity. The problem here corresponds more to the
    question which metric is the best one to solve a task, e.g. the dissimilarity of
    objects to. This has nothing to do with the choice of units of measurement (which
    in our case are [log$\_$2( p)] = bits).

    But the inappropriately chosen example reminds us of another point, which is
    extremely important and yet very elementary! If you add or subtract two
    quantities, they must have the same unit of measurement; otherwise you get
    complete nonsense. In our case we calculate AC - PD. AC has, because of log$\_$2 (p
    ), the unit bit; if we would calculate with log$\_$256 (p ) instead, we would have
    the result in the unit byte. So we have to calculate AC also in the unit "bit".
    Otherwise we pretend to be quantitative and yet we are only doing pseudo-science.

    This means that PD must also be an entropic quantity calculated with logarithms to
    base 2. This condition is fulfilled by the KL-divergence, Jensen-Shannon
    divergence and the conditional entropy. The condition is certainly not fulfilled
    for the Minkowski distances. The latter should therefore not be brought into play
    at all !

    \textcolor{orange}{I am also skeptical about the quantity D$\land$k$\_$new in Eq. (9), except for the case
    k=1.
    If one does a dimensional consideration by (for the moment) leaving out the
    summand 1 in the logarithm in Eq. (9),} then one can consider the exponent k as a
    multiplicative factor before the logarithm and thus before the sum; i.e. k is a
    factor that changes the unit of measurement. Although this is not a proof
    (especially since you cannot ignore the presence of the summand 1 in the
    logarithm), it is an indication that k unequal to 1 could be problematic. It could
    mean, so to say, to subtract bytes from bits.

    Aren't in information theory only terms of the type p * log (p ) considered, where
    p are always probabilities with values from the interval [0,1] and the logarithms
    are always taken to the same basis. I think there is a deep reason for this.

    I acknowledge the effort to make the selection of the "best" quantity halfway
    rational. But even the use of multi-criteria decision analysis (MCDA) cannot hide
    the fact that there is a great deal of arbitrariness in the analysis, as most
    numerical values were simply assumed. That these are somehow plausible is not
    enough to convince me. I think Sect. 4 is a rendition of the path the authors have
    taken, but not the way it should be presented. My recommendation would be to
    shorten this drastically.

    However, this does not dispel my fundamental reservations about the approach to
    finding an appropriate measure of PD. The examples shown with the tiny alphabets
    are, compared to realistic tasks, extremely simple and in my opinion not
    convincing. I would find it more convincing if, for example, 2 really realistic
    examples with realistic questions/tasks and suitable visualizations were used to
    find out which PD measure is most suitable.

    In doing so, I would leave out all candidates for whom it is actually clear from
    the outset that they will not perform well (e.g. for dimensionality reasons).

    In Section 5, the examples chosen again are too artificial and academic. The
    subway maps were primarily designed to quickly convey the number of stops between
    two locations (the most important information, e.g. to get off at the right
    station), and only secondly to estimate travel times (at least within the city,
    the travel time is approximately defined by the number of stops) and only thirdly
    to estimate actual distances.

    I also found the tomography example unworldly. Better would be e.g. the artery
    data set, but with reasonable questions. For example, a doctor would look for
    constrictions in the arteries, the spatial location of the constrictions in the
    arterial network; furthermore, she/he would estimate how narrow the constrictions
    really are (taking into account possible measurement and reconstruction errors)
    and what the medical risk is -- taking into account the overall blood flow in the
    arterial network and considering, which body regions are supplied.

    Overall, it is not quite clear to me why the motivating factor for this work,
    namely the boundedness of PD, is so important. Furthermore, I find the choice of a
    more suitable mathematical quantity not convincing. This selection should be made
    on the basis of real, realistic problems with appropriately designed
    visualizations. A summarizing discussion of the assumptions, limitations and scope
    of the theory is missing.

  Summary Rating

    <b>Reject</b>\\
    The paper is not ready for publication in SciVis / TVCG.\\
    The work may have some value but the paper requires major revisions or
    additional work that are beyond the scope of the conference review cycle to meet
    the quality standard. Without this I am not going to be able to return a score of
    '4 - Accept'.

  The Summary Review

    The authors aim to contribute to the foundation of data visualization by
    mathematically capturing essential characteristics of the visualization process.
    The main goal is a general mathematical expression that can be used to calculate
    the cost-benefit ratio of a visualization on an information-theoretical basis.
    While the costs are given by e.g. mean response times of the users, the benefit is
    more difficult to define. In Ref. [11], to which the first author of the current
    paper already contributed, a measure for the "benefit" was developed (Eq. 2). It
    consists of the difference "Alphabet Compression (AD)" - "Potential Distortion
    (PD)". The subject of the present work is PD, which measures the informative
    divergence between viewing the data through visualization with information loss
    and reading the data without any information loss. In [11] an expression based on
    Kullback-Leibler (KL) divergence was used for PD. Since the KL divergence is
    unbounded, PD can be arbitrarily large. In the eyes of the authors this is a
    problem which they - and this is the actual aim of the paper - want to solve by a
    more suitable mathematical expression.

    To this end, the authors first present a set of candidates for limited measures,
    oriented on common measures of information theory; some of these are
    parameterized, resulting in greater diversity. To select the most suitable
    measure, they define a set of criteria, apply them, using also multi-criteria
    decision analysis (MCDA). Then they validate the results with 2 case studies and
    conclude that a newly introduced measure, D$\land 2\_$new, is the most suitable one.

    All 4 reviewers assume that information theory will be part of the a foundational
    theory of visualization and will at least provide constraints for further
    approaches. \textcolor{teal}{However, the majority of the reviewers have doubts about the approach
    [11]. Since [11] and the follow-up work were positively assessed in peer reviews,
    these reviewers tried NOT to include their reservations in the evaluation, but to
    assess the manuscript on the basis of the assumption that the approach [11] is
    meaningful.}

    The main results of the review are:

    + The effort to make a contribution to the theoretical basis of the visualization
    is to be evaluated positively, especially since fundamental concepts are not
    sufficiently clarified and it is therefore a very difficult undertaking. (R1, R2,
    R3, R4)

    \textcolor{orange}{- The need for a bounded measure for PD is not sufficiently justified; there are
    several arguments (see in particular R2 and R3) that this requirement is not
    necessary. This, however, makes the premise of the paper doubtful. (R1, R3, R4)}

    \textcolor{orange}{- Some of the authors' trains of thought cannot be fully comprehended, as there
    are discrepancies with earlier work that are not being clarified. (R1)}

    \textcolor{orange}{- The alleged proofs are unsustainable. (R1, R3, R4)}

    \textcolor{orange}{- The proposed alternative measures are sometimes very ad hoc (R1, R3, R4), some
    are dimensionally doubtful (R4), they are (probably) not additive (R3) and there
    is no natural language interpretation (R1); other reasonable choices would have
    been possible (R3).}

    \textcolor{orange}{- The evaluation of measures has a strong ad hoc character (R1, R2, R3, R4); with
    the combination of Likert scales and MCDA almost any connection to information
    theory is lost (R3).}

    \textcolor{orange}{- The paper demands a lot from the reader; without detailed knowledge of
    information theory concepts, as well as details of the cost-benefit ratio, the
    paper is difficult to understand. It would be better if the basic terms and
    concepts were briefly presented again in the paper, with reference to the
    appendix) (R1). It would be nice if some of the two case studies were interwoven
    into the mathematical presentation to better anchor the concepts (R2). The
    notation needs to be improved (R1),}

    \textcolor{orange}{- The validation of the PD measure is questionable (R1, R3, R4); for details see
    in particular R3.}

   \textcolor{violet}{- When trying to apply the presented theory to real applications, huge gaps open
    up (R1, R3, R4); it starts with the fact that it is the goal of almost all
    visualizations to reduce the amount of information and to transmit only the
    information that is essential for the respective application situation; and it
    ends with the fact that the case studies in the paper, on closer inspection,
    hardly support the goals of the work (R1, R3, R4).}

    \textcolor{violet}{- A summarizing discussion of the assumptions, limitations and scope of the theory
    presented is missing (R1, R4).}

----------------------------------------------------------------

\end{narrowfont}
\normalfont
\normalsize
\setlength{\parindent}{5.1mm}
\setlength{\parskip}{0pt}

\end{document}